\documentclass[twocolumn,tighten,times]{aastex701}
\usepackage{amsmath}
\usepackage{multirow}

\newcommand{\rtr}{\ensuremath{R_\text{tr}}}
\newcommand{\vtr}{\ensuremath{v_\text{tr}}}
\newcommand{\mej}{\ensuremath{M_\text{ej}}}
\newcommand{\tej}{\ensuremath{T_\text{ej}}}
\newcommand{\teff}{\ensuremath{T_\text{eff}}}
\newcommand{\tout}{\ensuremath{T_\text{out}}}
\newcommand{\tejin}{\ensuremath{T_\text{ej,in}}}
\newcommand{\mcsm}{\ensuremath{M_\text{CBM}}}
\newcommand{\rcsm}{\ensuremath{R_\text{CBM}}}
\newcommand{\msun}{\ensuremath{M_\odot}}
\newcommand{\rhoej}{\ensuremath{\rho_\text{ej}}}
\newcommand{\rhocsm}{\ensuremath{\rho_\text{CBM}}}
\newcommand{\rhowind}{\ensuremath{\rho_\text{wind}}}
\newcommand{\fomega}{\ensuremath{f_\Omega}}
\newcommand{\mdot}{\ensuremath{\dot{M}}}
\newcommand{\ergss}{erg\,s$^{-1}$}
\newcommand{\kms}{km\,s$^{-1}$}
\newcommand{\xion}{\ensuremath{x_\text{ion}}}
\newcommand{\lbol}{\ensuremath{L_\text{bol}}}

\shorttitle{2D simulations of Luminous Red Novae }
\shortauthors{Kirilov et al.}

\begin{document}
\title{Two-Dimensional Radiation-Hydrodynamic Simulations of Luminous Red Novae}

\author[orcid=0000-0002-1564-4920,sname=Kirilov,gname=Anthony]{Anthony Kirilov}
\affiliation{Institute of Theoretical Physics, Faculty of Mathematics and Physics, Charles University, Prague, Czechia}
\email{anthony.kirilov@matfyz.cuni.cz}

\author[orcid=0000-0002-9019-9951,gname=Diego, sname=Calder\'{o}n]{Diego Calder\'{o}n} 
\altaffiliation{Alexander von Humboldt Fellow}
\affiliation{Max-Planck-Institut für Astrophysik, Karl-Schwarzschild-Straße 1, 85748 Garching, Germany}
\email{calderon@mpa-garching.mpg.de}

\author[orcid=0000-0003-2512-2170,gname=Ond\v{r}ej,sname=Pejcha]{Ond\v{r}ej Pejcha}
\affiliation{Institute of Theoretical Physics, Faculty of Mathematics and Physics, Charles University, Prague, Czechia}
\email[show]{ondrej.pejcha@matfyz.cuni.cz}

\author[orcid=0000-0001-7626-9629,sname=Duffell,gname=Paul]{Paul C. Duffell}
\affiliation{Department of Physics and Astronomy, Purdue University, 525 Northwestern Avenue, West Lafayette, IN 47907, US}
\email{pduffell@purdue.edu}

\correspondingauthor{Ond\v{r}ej Pejcha}

\begin{abstract}
Luminous Red Novae (LRNe) are transients associated with mass ejection during stellar mergers and common envelope evolution (CEE). LRNe have the potential to illuminate the poorly understood phases of binary evolution leading up to the CEE, during the mass ejection phase, and in the immediate aftermath. However, the mechanism responsible for powering LRN light curves and the origin of their observed diversity remain open questions. Here, we perform two-dimensional moving-mesh radiation-hydrodynamic simulations of LRNe that take into account hydrogen and helium recombination and relevant opacities. We study a typical high-mass stellar merger, which dynamically ejects $2\,\msun$ with a characteristic velocity of $410$\,\kms. This ejecta collides with $2.7\,\msun$ of equatorially concentrated circumbinary material (CBM) left behind from a prior phase of non-conservative runaway mass transfer. We find that the resulting light curve is composed of a short, blue peak followed by a redder, predominantly shock-powered plateau with luminosities reaching up to $10^{41}$\,\ergss\ and durations up to 200\,days. These luminosities are significantly higher, and the durations much longer, than those produced by a simple spherical ejection of the same mass. They also depend in a complex way on the radial distribution of the CBM and the viewing angle. The shock is embedded in the ejecta and its observational signatures during the optically-thick phase are largely hidden. Our results are broadly compatible with observations of the brightest extragalactic LRNe and pave the way for the transformation of LRNe into powerful probes of binary evolution.
\end{abstract}



\section{Introduction}
Many evolutionary pathways of binary stars critically rely on a short, dramatic evolutionary phase, common envelope evolution (CEE) \citep{Paczynski76}. During CEE, the binary loses substantial amounts of its mass, angular momentum, and energy, which facilitates the formation of systems such as compact object binaries composed of black holes, neutron stars and white
dwarfs, cataclysmic variables, X-ray binaries, and gravitational wave sources. Alternatively, with significantly less mass loss, the two stars coalesce into a single, potentially exotic object. Due to a wide range of physical and temporal scales, as well as the variety of potentially important astrophysical processes, the theoretical understanding of CEE remains insufficient, causing significant uncertainties in the interpretation of observations \citep[e.g.][]{Ivanova2013_review,iaconi18,roepke2023,gagnier25}.

The dynamical phase of CEE has been associated with a class of transient brightenings called Luminous Red Novae (LRNe) \citep{Soker03,Soker2006,tylenda06,Tylenda2011,Ivanova2013_transient}. LRNe reach peak luminosities between about $10^{37}$ and few times $10^{41}$\,\ergss, their expansion velocities range between about $100$ and $1000$\,\kms,  they consistently evolve to low temperatures and often show evidence of asymmetries as well as dust and molecule formation \citep[e.g.,][]{banerjee03,tylenda05,blagorodnova17,woodward21}. The light curves of LRNe typically show multiyear slow brightening from the quiescent level followed by a fast brightening to the first peak, which is often followed by an extended plateau or a second peak \citep[e.g.,][]{Tylenda2011,kankare15,blagorodnova20}. The typical time interval between the first peak and the end of the second peak or plateau ranges from approximately 10 to 200 days \citep[e.g.][]{Pastorello2019,Blagorodnova2021}.

LRNe and their remnants hold great promise for illuminating the poorly understood physics of CEE, however, several conceptual open problems and differences with respect to better understood transients such as Type II supernovae have prevented analysis of LRNe from reaching this potential. First, most multidimensional CEE simulations study (super)giant primaries with close binaries as the likely outcome, but LRN progenitors appear significantly less expanded and likely leave behind merged objects \citep{Blagorodnova2021,Klencki2021}. Second, the vast majority of multidimensional CEE simulations performed so far did not self-consistently include radiation transport, which makes their utility in explaining LRNe limited \citep[with the exception of][]{ricker19,hatfull25,lau25}. Both of these points motivate the development of theoretical models specifically for LRNe.

Finally and most importantly, it is not yet clear what powers LRN light curves. Plateau-shaped light curves are most naturally produced by energy diffusion from expanding hydrogen-rich envelopes ejected and energized during the binary interaction \citep{Ivanova2013_transient,macleod17}. Unlike Type IIP supernovae, slower expansion velocities and higher densities in LRNe make the ejected envelopes not simply radiation-dominated and instead effects such as hydrogen recombination energy, $E_\text{rec} \approx 2\times 10^{46} (\mej/\msun)$\,erg, can appreciably contribute to the light curve when released over several months. Early hydrodynamical models along this direction were calculated by \citet{lipunov17}. More recently, \citet{Chen24} successfully modeled the light curve of AT2019zhd \citep{Pastorello2021_2019zhd} that peaked at $2\times 10^{39}$\,\ergss\ by injecting a time-variable outflow at the inner boundary of their spherically symmetric radiation hydrodynamic model with flux-limited diffusion and realistic equation of state and opacities. However, scaling relations developed specifically for LRNe suggest that explaining bright events with $\sim 10^{41}$\,\ergss\ and $\sim 200$-day long plateaus requires unrealistically high ejecta masses \citep{Matsumoto2022}. Other theories explain LRNe by the action of jets launched by a star spiralling in the common envelope \citep[e.g.,][]{soker16,soker2020,soker2023}. Furthermore, the effect of irradiation by the central remnant of the merger, which should remain relatively bright on timescales longer than the duration of LRNe \citep{schneider19}, remains unexplored.

\citet{Metzger2017} proposed an alternative mechanism for LRNs based on a shock interaction between the roughly spherical dynamically ejected material and the preexisting equatorially concentrated circumbinary mass distribution (CBM). The first peak is caused by cooling emission from the faster ejecta, while the embedded and reprocessed shock power can lead to secondary peaks or plateaus about ten times brighter and twice as long than what is reasonably possible in the recombination model \citep{matsumoto25}. The radiative processes accompanying the accumulation of CBM before the merger can explain the gradual slow brightening of LRN progenitors \citep{pejcha14,pejcha16_bound,pejcha16_cool,Pejcha2017,tsuna24} and naturally connect LRNe to the strongly non-conservative runaway mass transfer seen in binary evolution models \citep[e.g.,][]{Blagorodnova2021,Klencki2021,Marchant2021}. Furthermore, the asymmetric shock interaction produces bipolar shapes similar to those observed in LRN remnants \citep[e.g.,][]{chesneau14,kaminski18,steinmetz24}. Although these features are attractive, the challenging multidimensional nature of the model has so far prevented more quantitative checks of its viability going beyond semianalytic estimates and 1D models. 

In this work, we present axisymmetric radiation hydrodynamic simulations of the first year of LRNe. Our goal is to demonstrate that shocks embedded in ejecta can power the brightest and longest LRNe under generic assumptions on the masses of the ejecta and the CBM and their thermodynamic structure. In Section~\ref{sec:setup}, we describe our simulations and initial conditions. In Section~\ref{sec:results}, we present evolution of the thermodynamic and radiation structure of the ejecta, light curves, evolution of the photospheric temperature, viewing-angle dependence, effect of central irradiation by the merger remnant, and perform a basic parameter study of the CBM properties along with comparison with observed events. In Section~\ref{sec:disc}, we summarize our findings and discuss implications for LRNe.

\section{Simulation setup}
\label{sec:setup}

\begin{deluxetable*}{llcccccccccc}
\tabletypesize{\scriptsize}
\tablewidth{0pt} 
\tablecaption{Parameters of simulation runs \label{tab:params}}
\tablehead{
\colhead{EOS, opacities} & \colhead{Run} & $\kappa_\text{R,floor}$ & \colhead{$\max \kappa_\text{P}/\kappa_\text{R}$}  &\colhead{$\mej$} &  \colhead{$E_\text{ej,kin}$} & \colhead{$\tejin$}     & \colhead{$\mdot_\text{wind}$}        & \colhead{$\mcsm$}       & \colhead{$\rcsm$}           & \colhead{$N_r$} & \colhead{$N_\theta$} \\
\colhead{}  &   \colhead{} & \colhead{cm$^2$\,g$^{-1}$}& \colhead{} & \colhead{($\msun$)}& \colhead{($10^{50}$\,erg)} & \colhead{($10^5$\,K)} & \colhead{($\msun$\,yr$^{-1}$)}  & \colhead{($\msun$)}  & \colhead{($10^{15}$\,cm)} & \colhead{}     & \colhead{}         
} 
\startdata
\multirow{21}{*}{Table} & \textbf{d0.1hiE} &  $10^{-3}$ &  \multirow{11}{*}{10} &\multirow{11}{*}{2.0}&  \multirow{11}{*}{0.02}& \multirow{11}{*}{5.0}  & $7\times 10^{-6}$ & 2.7 & 0.1 & \multirow{11}{*}{512} & \multirow{11}{*}{64}\\
& \textbf{d0.3hiE}                         & $10^{-3}$& & & & & $7\times 10^{-6}$ & 2.7 & 0.3 & &  \\
& \textbf{d2.0hiE}                         & $10^{-3}$& & & & &$7\times 10^{-6}$ & 2.7 & 2.0 &  &  \\
& e0.3hiE                         & $10^{-3}$& & & & & $7\times 10^{-5}$ & 2.7 & 0.3 & &  \\
& \textbf{enocbmhiE}                       & $10^{-3}$ & & & & & $7\times 10^{-5}$ & 0.0 & -- & &  \\
& g0.3hiE\tablenotemark{b}                         & $10^{-4}$& & & & & $7\times 10^{-5}$ & 2.7 & 0.3 & &  \\
& h0.3hiE\tablenotemark{b}                         & $10^{-5}$& & & & & $7\times 10^{-5}$ & 2.7 & 0.3 & &  \\
& n0.1hiE\tablenotemark{b}                      & 0& & & & & $7\times 10^{-5}$ & 2.7 & 0.1 & &  \\
& n0.3hiE                         & 0& & & & & $7\times 10^{-5}$ & 2.7 & 0.3 & &  \\
& n2.0hiE\tablenotemark{b}        & 0& & & & & $7\times 10^{-5}$ & 2.7 & 2.0 & &  \\
& nnocbmhiE         & 0& & & & & $7\times 10^{-5}$ & 0.0 & -- & &  \\
\cline{2-12}
& \textbf{d0.1loE}                        &  $10^{-3}$ &  \multirow{9}{*}{10} & \multirow{9}{*}{2.0}&  \multirow{9}{*}{0.02}& \multirow{9}{*}{1.0}  & $7\times 10^{-6}$ & 2.7 & 0.1 & \multirow{9}{*}{512} & \multirow{9}{*}{64} \\
& \textbf{d0.3loE}                         & $10^{-3}$& & & &  & $7\times 10^{-6}$ & 2.7 & 0.3 &  &  \\
& \textbf{d2.0loE}                         & $10^{-3}$& & & & & $7\times 10^{-6}$ & 2.7 & 2.0 &  &  \\
& e0.3loE                                  & $10^{-3}$& & & &  & $7\times 10^{-5}$ & 2.7 & 0.3 &  &  \\
& \textbf{enocbmloE}                       & $10^{-3}$ & & &   & & {$7\times 10^{-5}$} & 0.0 & -- &  &  \\
& n0.1loE\tablenotemark{b}                      & 0& & & & & $7\times 10^{-5}$ & 2.7 & 0.1 & &  \\
& n0.3loE\tablenotemark{b}                         & 0& & & & & $7\times 10^{-5}$ & 2.7 & 0.3 & &  \\
& n2.0loE\tablenotemark{b}        & 0& & & & & $7\times 10^{-5}$ & 2.7 & 2.0 & &  \\
& nnocbmloE\tablenotemark{b}         & 0& & & & & $7\times 10^{-5}$ & 0.0 & -- & &  \\
\cline{2-12}
& nIIP3e5                      &0 & 10 & 10.0&  9.8 & 3  & $7\times 10^{-5}$ & 0.0 & -- & 512 & 64 \\
\hline
\multirow{15}{*}{H only, analytic} & s0.1hiE                        &\multirow{4}{*}{$2\times 10^{-3}$} & \multirow{4}{*}{1} &  \multirow{4}{*}{2.0}&  \multirow{4}{*}{0.02}& \multirow{4}{*}{5.0}  & \multirow{4}{*}{$7\times 10^{-8}$} & 2.7 & 0.1 & \multirow{4}{*}{512} & \multirow{4}{*}{64} \\
& s0.3hiE                        & & & &  & &  & 2.7 & 0.3 &  &  \\
& s2.0hiE                        & & & &  & & & 2.7 & 2.0 &  &  \\
& snocbmhiE                      & & & & & &  & 0.0 & -- &  &  \\
\cline{2-12}
& s0.1loE                         &\multirow{6}{*}{$2\times 10^{-3}$} & \multirow{6}{*}{1} &  \multirow{6}{*}{2.0}&  \multirow{6}{*}{0.02}& \multirow{6}{*}{1.0}  & $7\times 10^{-8}$ & 2.7 & 0.1 & 512 & \multirow{6}{*}{64} \\
& \textbf{s0.1loEirr}\tablenotemark{c}                         & & & & &  & {$7\times 10^{-7}$} & 2.7 & 0.1 & 512 &  \\
& s0.3loE                         & & & & &  & {$7\times 10^{-8}$} & 2.7 & 0.3 & 512 &  \\
& s2.0loE                        & &  & & & & {$7\times 10^{-8}$} & 2.7 & 2.0 & 512 &  \\
& slowTout\tablenotemark{a}                     & &   & & & & {$7\times 10^{-8}$} & 2.7   &  2.0 & 1024 &  \\
& snocbmloE                      & &  & &  & & {$7\times 10^{-8}$} & 0.0 & -- & 512 &  \\
\cline{2-12}
& smedr                         &\multirow{4}{*}{$2\times 10^{-3}$} & \multirow{4}{*}{1} & \multirow{4}{*}{2.0}&  \multirow{4}{*}{0.02} & \multirow{4}{*}{1.0}  & \multirow{4}{*}{$7\times 10^{-8}$} & \multirow{4}{*}{2.7} & \multirow{4}{*}{2.0} & 1024 & 64 \\
& ssuperr                        & & & & & &  &  &  & 2048 & 64 \\
& smedth                         & & & & &  &  &  &  & 512 & 128 \\
& shires                         & & & & & &  & &  & 1024 & 128 \\
\cline{2-12}
& IIP2e5                      &\multirow{2}{*}{$2\times 10^{-3}$} & \multirow{2}{*}{1} & \multirow{2}{*}{10.0}&  \multirow{2}{*}{9.8} & 2  & \multirow{2}{*}{$7\times 10^{-8}$} & \multirow{2}{*}{0.0} & \multirow{2}{*}{--} & \multirow{2}{*}{512} & \multirow{2}{*}{64} \\
& IIP4e5                      & & & & & 4  &  &  &  &  &  \\
\hline
\multirow{6}{*}{$\Gamma$-law, analytic} & s0.3hiEgamma   &\multirow{2}{*}{$2\times 10^{-3}$} &\multirow{2}{*}{1}&  \multirow{2}{*}{2.0} & \multirow{2}{*}{0.02} & \multirow{2}{*}{5.0} & \multirow{2}{*}{$7\times 10^{-8}$} & 2.7 & 0.3 & 1024 &  \multirow{2}{*}{64} \\
& snocbmhiEgamma   & & &  &  & & & 0.0 & -- & 512 &  \\
\cline{2-12}
& s0.3loEgamma    &\multirow{2}{*}{$2\times 10^{-3}$} & \multirow{2}{*}{1} & \multirow{2}{*}{2.0} & \multirow{2}{*}{0.02} & \multirow{2}{*}{1.0} & \multirow{2}{*}{$7\times 10^{-8}$} & 2.7 & 0.3 & 1024 & \multirow{2}{*}{64} \\
& snocbmloEgamma & & & & & &  & 0.0 & -- & 512 &  \\
\cline{2-12}
& IIP2e5gamma  &\multirow{2}{*}{$2\times 10^{-3}$} & \multirow{2}{*}{1}& \multirow{2}{*}{10.0} & \multirow{2}{*}{9.8} & 2  & \multirow{2}{*}{$7\times 10^{-8}$} & \multirow{2}{*}{0.0} & -- & \multirow{2}{*}{512} & \multirow{2}{*}{64} \\
& IIP4e5gamma   & & &  & & 4  &  &  & -- &  &  \\
\enddata
\tablecomments{Most of the analysis is based on runs highlighted in bold.}
\tablenotetext{a}{Test run with significantly lower $T_\text{out}$.}
\tablenotetext{b}{Simulation did not reach the post-plateau phase due to numerical difficulties.}
\tablenotetext{c}{Includes irradiation by the central source.}
\end{deluxetable*}

We perform two-dimensional axisymmetric $(r,\theta)$ radiation-hydrodynamic simulations using the code RJET. This code is built on top of JET \citep{Duffell11,Duffell2013}, which is a finite-volume hydrodynamic code using a spherical mesh with each radial wedge moving radially with the flow, making it ideal for modeling explosive events. \citet{Calderon2021,Caldron2024} developed a radiation treatment and coupling module for JET based on the mixed-frame formulation \citep{Krumholz2007} under the gray flux-limited diffusion (FLD) approximation \citep{alme1973}. Radiation quantities were linearized and updated implicitly, alleviating the constraints on the timestep. In this work, we further developed RJET by implementing a non-linear Newton-Raphson solver for radiation quantities, general equation of state (EOS) and opacities, and modifications to timestep and mesh (de-)refinement controls. Specifically, for LRNe, we use an EOS that takes into account hydrogen and helium ionization and recombination as well as molecular hydrogen. We use analytic and tabulated opacities that incorporate Kramers, Thomson scattering, H$^-$, molecules, and dust. Details of the code, basic checks, validation, and resolution study are given in Appendix~\ref{app:code}. In the rest of this Section, we discuss initial conditions for our simulations.

\subsection{Initial and boundary conditions}

Rather than attempting to model specific events, we are interested in exploring whether the shock-powered model can reasonably explain the brightest observed LRNe. Inspired by AT2018bwo \citep{Blagorodnova2021}, other progenitor mass determinations \citep{cai22}, and theoretical models \citep[e.g.,][]{klencki25}, we assume that LRNe arise from binaries with a total mass of about $M_*=20\,\msun$ and a characteristic size of about $a=10^{13}$\,cm. The initial conditions are composed of dynamical merger ejecta, pre-existing equatorially-concentrated CBM distribution, and a background wind-like density, which is present mostly for numeric reasons. We also include the possibility of a central radiation source with luminosity $L_\text{irr}$.

We assume that the dynamical phase of the merger ejects $\mej = 2\,\msun$, which is compatible with simulations of stellar mergers showing that up to about $10\%$ of the mass of the binary is ejected \citep{lombardi02,glebbeek13}. Since we did not simulate the merger itself, we do not know the density and thermal structure of the ejecta. This is a significant complication with respect to studies of core-collapse supernovae, where an energetic thermal or momentum bomb placed at the center of a progenitor star produces a well-defined ejecta profile. In stellar mergers, the ejecta expand close to the escape velocity from the binary, and the details of the ejection process matter. Consequently, inspired by semi-analytic studies of supernovae \citep{Chevalier1989,mcdowell18,kurfurst20}, we decided to begin with homologous expansion and a broken power-law density profile,
\begin{equation}
    \rhoej(r) = \frac{(n-3)(3-d)}{4\pi(n-d)} \frac{\mej}{\rtr^3}
    \begin{cases}
        \left(r/\rtr\right)^{-d} & \text{$r \leq \rtr$},\\
        \left(r/\rtr\right)^{-n} & \text{$r > \rtr$},
    \end{cases} 
    \label{eq:rhoej}
\end{equation}
where $\rtr = \vtr t$  and $\vtr$ is a characteristic velocity defined as
\begin{equation}
\vtr^{2} =\frac{2(5-d)(n-5)}{(3-d)(n-3)}\frac{E_\text{ej,kin}}{M_\text{ej}}.
\label{eq:vtr}
\end{equation}
We choose $E_\text{ej,kin} = 2\times 10^{48}$\,erg, $d=1$, and $n=16$, which gives $\vtr \approx 410\,$\,\kms, above the escape velocity of the binary, $v_\text{esc} = \sqrt{2GM_*/a} \approx 230$\,\kms. At the start of our simulations, $t_0 = 3.1$\,days, $\rtr \approx 1.1 \times 10^{13}$\,cm.  We set the initial ejecta temperature profile to
\begin{equation}
    \tej(r) = \begin{cases}
    \tejin & r \leq \rtr,\\
    \max \left[\tout, \tejin\left(\frac{\rhoej(r)}{\rhoej(\rtr)}\right)^{1/3}\right] & r > \rtr.\\
    \end{cases}
    \label{eq:temp}
\end{equation}
In parameterized initial conditions, the initial thermal content of the ejecta should not be significantly higher than $E_\text{ej,kin}$ so that the expansion properties are not significantly influenced by $P\text{d}V$ work. To satisfy this constraint and describe two different thermodynamical states of the ejecta, we choose two different values of $\tejin = 5\times 10^5$\,K and $1\times 10^5$\,K.  At $t_0$, $\tejin = 5\times 10^5$\,K makes the ejecta radiation-dominated with $E_\text{ej,rad} \approx 3.1\times 10^{48}$\,erg. For $\tejin = 1\times 10^5$\,K, the ejecta is dominated by an ideal gas with $E_\text{ej,gas} = (3/2)(k_\text{B}/m_\text{p}) \mej T \approx 7.5\times 10^{46}$\,erg. In both cases, the ejecta also contains about $4\times 10^{46}$\,erg of hydrogen recombination energy. As we show in Appendix~\ref{app:tests}, the effect of helium recombination is minimal.

We assume that the dynamical merger was preceded by a phase of runaway mass transfer between the binary components. This mass transfer was highly non-conservative, and a significant part of the transferred mass left the binary in the vicinity of the L2 point. Depending on binary properties, this material will be organized in an equatorially-concentrated outflow or disk, possibly surrounded by disk winds or jets \citep{shu79,pejcha16_cool,pejcha16_bound,hubova19,shiber19,Lu2022}. The density profile is specific to a particular binary and its history of mass loss. For simplicity, we choose an exponentially cut CBM density profile that extends $0.1\pi$ on each side of the equator and covers the solid angle fraction $f_\Omega = 0.3$, 
\begin{equation}
\rhocsm(r) = \frac{\mcsm}{4\pi\fomega \rcsm r^2} \frac{2}{\sqrt{\pi}}\exp\left[ - \left(\frac{r}{\rcsm} \right)^2 \right]
\label{eq:rhocsm}
\end{equation}
The CBM is blended with the ejecta at around $3\rtr$, roughly corresponding to an inner edge of a circumbinary disk located at $\sim 3a$ \citep{shu79} and ensures that the shock collision begins in an optically thin region of the ejecta.
We set $\mcsm = 2.7\,\msun$ to be approximately compatible with pre-merger binary evolution models \citep{Blagorodnova2021} and vary $\rcsm$ between $0.1 \times 10^{15}$\,cm and $2.0\times 10^{15}$\,cm to explore what happens when the shock encounters the exponential cutoff within or after the optically thick phase. Since $\rhocsm$ is blended with $\rhoej$ at some nonzero radius, the CBM mass in the simulations is not exactly equal to $\mcsm$ and slightly depends on $\rcsm$. We set the CBM velocity to $110$\,\kms and its initial temperature to 1500\,K.

Finally, for numerical reasons, we add a background wind-like density profile
\begin{equation}
    \rhowind(r) = \frac{\mdot_\text{wind}}{4\pi v_\text{wind} r^{2}}.
    \label{eq:rhowind}
\end{equation}
with $v_\text{wind} = 110$\,\kms. We vary $\mdot_\text{wind}$ between $7\times 10^{-8}$ and $7\times 10^{-5}$\,\msun\,yr$^{-1}$. These values are sufficiently small to ensure that collision with the dynamical ejecta does not appreciably contribute to the total luminosity. We set the wind and CBM temperatures equal to $\tout$.

We blend all three components in radius and angle with a simple Fermi-Dirac function, ensuring that the transitions are sufficiently resolved on our grid. Initially, gas and radiation are in equilibrium everywhere. Irradiation by the central source is implemented as an inner boundary condition considered in the radiation implicit step in the same manner as in \cite{Calderon2021,Caldron2024}.
Our grid initially extends between $10^{12}$ and $4\times 10^{13}$\,cm radially and between $[0,\pi/2]$ along the polar direction. Zones are spaced linearly in angle and initially logarithmically in radius. Most of our simulations start with $N_\theta = 64$ angular and $N_r = 512$ radial zones. The number of radial zones changes during the simulation due to mesh (de-)refinement while the number of angular zones and their spacing remains constant. Additionally, we run several models with higher resolution. 

As the grid expands, we need to provide thermodynamic and kinematic properties of the newly included regions. There is some uncertainty in this process, because radiation can influence these regions through deposition of energy and momentum. For densities and radial velocities we use the initial condition prescriptions described above. We set tangential velocity of the newly included regions to zero. The stability of the implicit solver is sensitive to the values of gas and radiation temperature, especially in the runs using tabulated EOS and opacities. In the newly  included zones, we set the gas temperature to be slightly below the next inner cell in radius and we set radiation energy density assuming constant radial optically-thin flux. These choices are made mostly for numerical reasons and to prevent artificial ingestion of energy from the outer boundary, but radiation is typically able to almost instantly bring the newly added cells to their physical values. In addition to varying $\tej$ and $\rcsm$, we calculate several models without CBM and with a simple $\Gamma$-law EOS. Our simulations typically end at $t=310$\,days. Our simulations are summarized in Table~\ref{tab:params}.

\section{Results}
\label{sec:results}

In this section, we present our results focusing on the hydrodynamic structure and its evolution (Sec.~\ref{sec:overall}), radiative structure (Sec.~\ref{sec:flux}), light curves (Sec.~\ref{sec:lc}), photospheric temperature (Sec.~\ref{sec:teff}), viewing angle dependence using raytracing (Sec.~\ref{sec:ray}), effect of the central luminous remnant (Sec.~\ref{sec:remnant}), and comparison to observed LRNe (Sec.~\ref{sec:comparison}).  In Appendix~\ref{app:movies}, we present animations showing the evolution of the main quantities analyzed in this Section for the six main runs with CBM. 

\subsection{Overall evolution}
\label{sec:overall}

\begin{figure*}
\plottwo{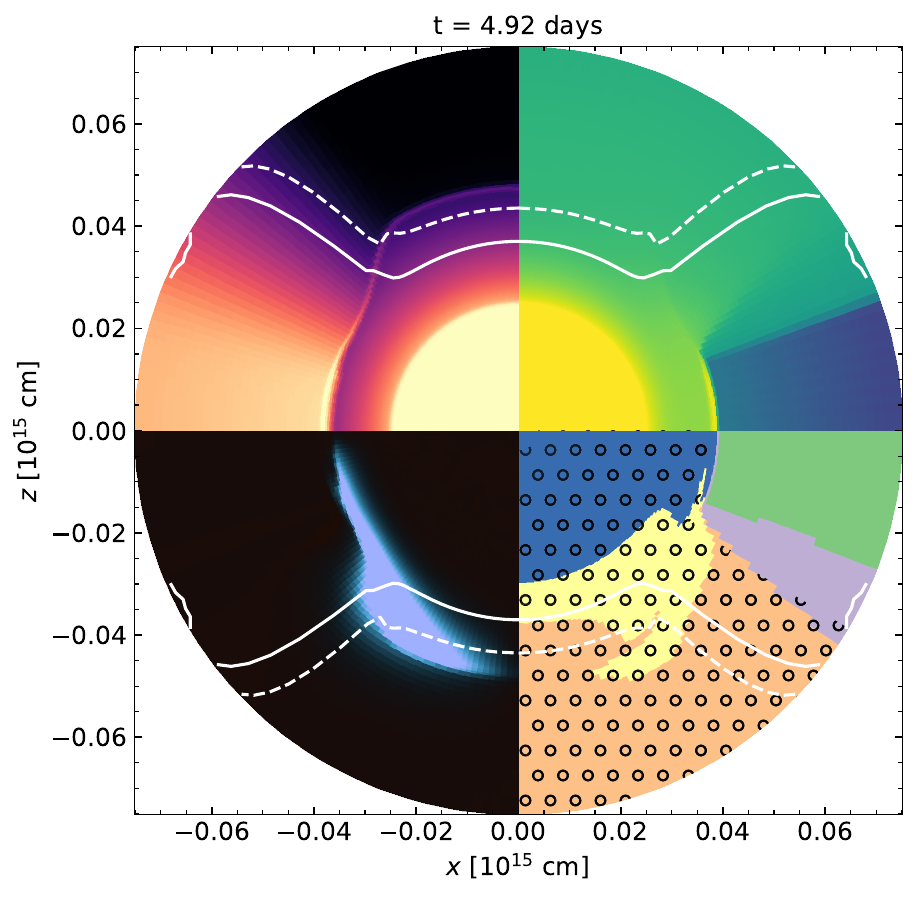}{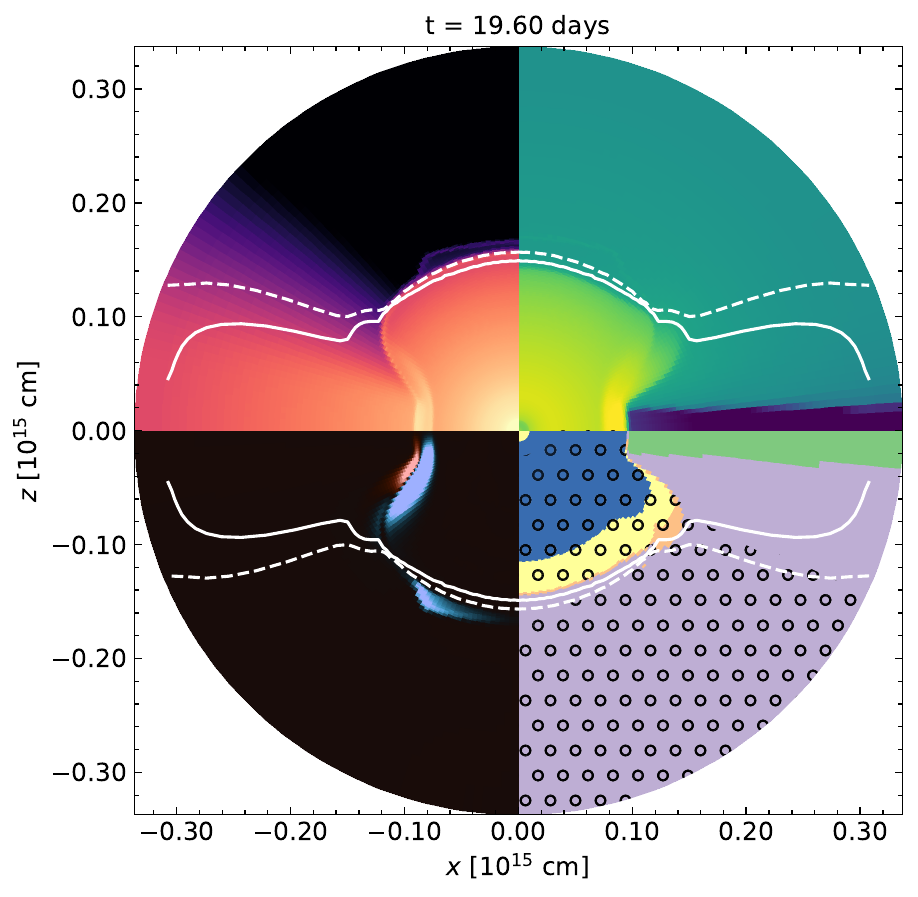}
\plottwo{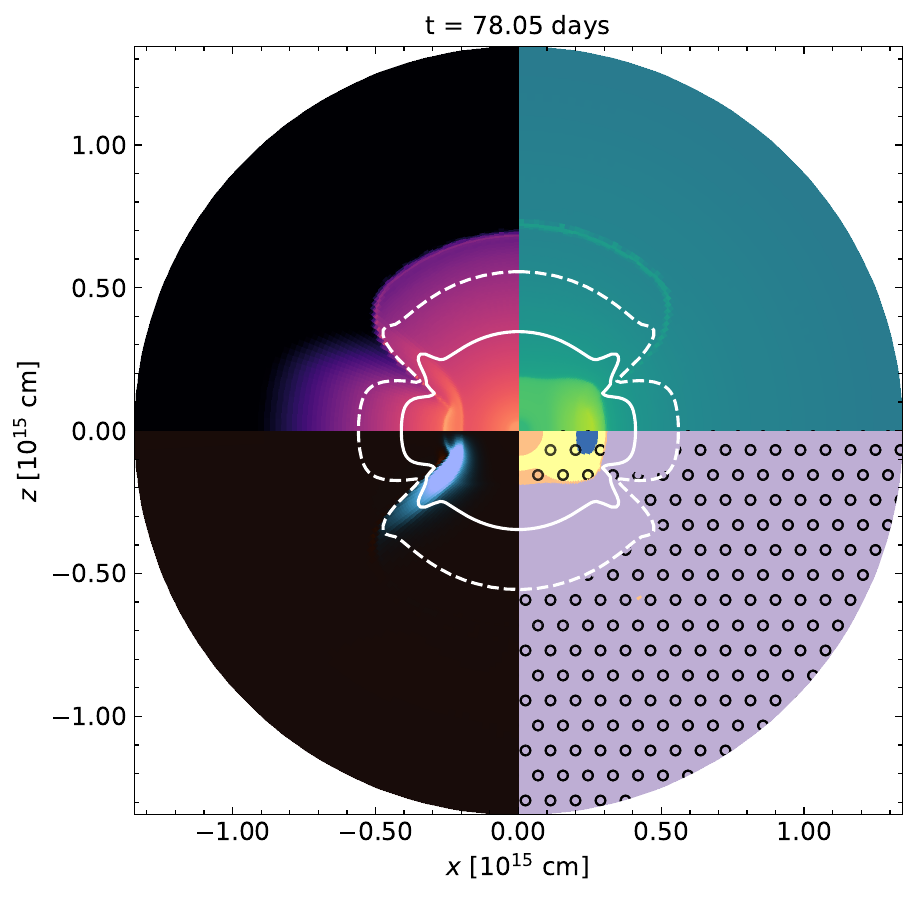}{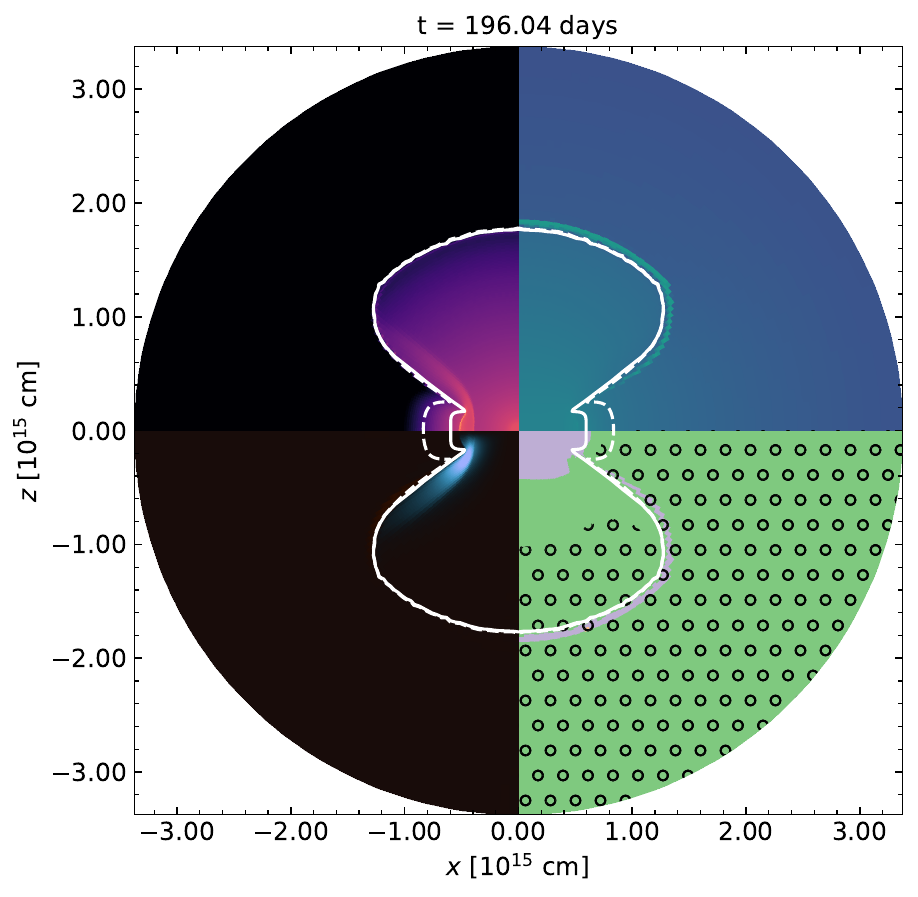}
\plotone{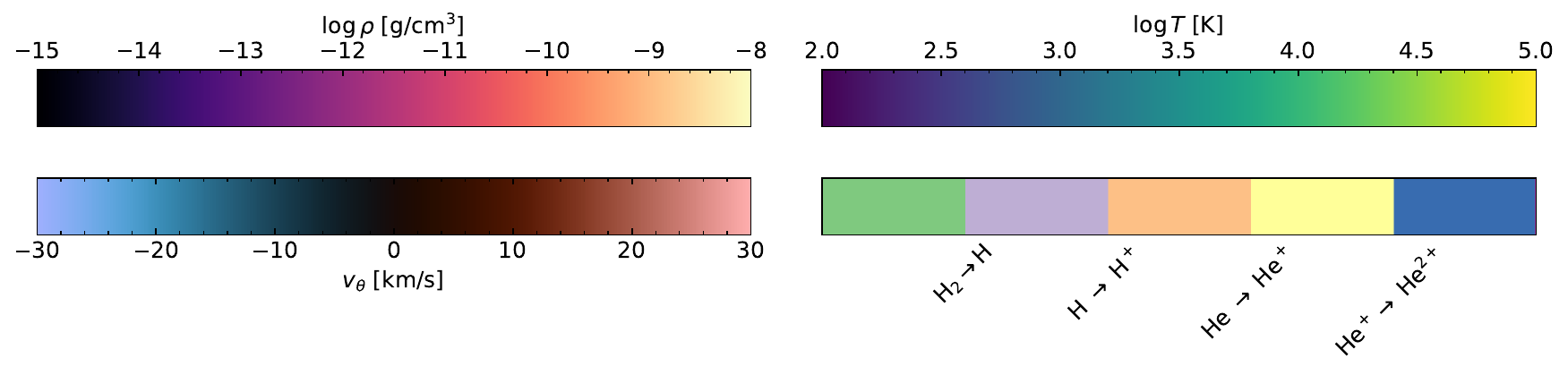}
\caption{Structure and evolution of the simulation d0.3hiE ($\rcsm=0.3 \times 10^{15}$\,cm, $\tejin = 5\times 10^5$\,K) at four representative epochs. Each quadrant of each panel shows different quantity: density $\rho$ (upper left), gas temperature $T$ (upper right), tangential velocity $v_\theta$ (lower left), and ionization structure (lower right). Explanation of the color scale scale is given in the legend at the bottom. Tangential velocity is positive for motions toward the plane of symmetry. Ionization structure is based on the evaluation of Saha-like equations for each species with more detailed explanation shown in Figure~\ref{fig:eos}. Hatched circles show regions where $a_\text{r}T^4/3$ exceeds gas pressure, but the dynamical effect of radiation depends on the local value of the flux-limiter (Eq.~[\ref{eq:app_mom}]).
White line shows the location of the $\tau=2/3$ surface measured along radial rays with solid and dashed lines showing results based on Rosseland and Planck mean opacities, respectively. \label{fig:overview}}
\end{figure*}

In Figure~\ref{fig:overview}, we show the structure and evolution of the density $\rho$, gas temperature $T$, tangential velocity $v_\theta$ and the ionization structure for run d0.3hiE. We will also compare to the low-$\tejin$ model d0.1loE. Animated version of Figure~\ref{fig:overview} and similar information for all CBM runs is available in Appendix~\ref{app:movies} in Figures~\ref{fig:d0.1hiE}--\ref{fig:d2.0loE}. The evolution begins by an almost immediate formation of a radiative shock at $r=3\times 10^{13}$\,cm due to the velocity difference between the ejecta and wind/CBM. The contrast in ejecta--wind density is so large that the shock at high latitudes does not significantly affect the ejecta expansion or the radiative properties, which we verify explicitly in Appendix~\ref{app:tests}, although it remains visible throughout the simulation as a region of locally increased $T$ and $\rho$. In equatorial regions, the radiation originating from the ejecta--CBM shock starts to influence its surroundings within less than a day after $t_0$. Since the collision initially happens in a low optical depth region, the shock radiation quickly ionizes the outer parts of the ejecta and the rest of the simulation domain except the dense equatorial CBM. For lower $\tejin$, the ionization of the outer ejecta layers is weaker. Simultaneously with the ionization, the radiation pressure pushes gas laterally away from the plane symmetry, as revealed by the negative values of $v_\theta$. Part of the hot gas is squeezed out from the shock regions at around $30\deg$ from the equatorial plane, but this feature likely depends on the vertical profile of the CBM. For higher $\tejin$, the initial thermal energy of the gas is able to somewhat affect the expansion of the ejecta, leading to several features visible within the first 5 days of the simulation (see Sec.~\ref{sec:flux} and \ref{sec:lc} and Figs.~\ref{fig:flux} and \ref{fig:lc} for more details). 

The declining shock power causes the region of complete ionization to retreat until it coincides with the photosphere, which happens at around 6 and 15 days for low and high $\tejin$ models, respectively. Afterward, the evolution proceeds smoothly until the reverse shock reaches the dense parts of the ejecta at $r<\rtr$ at around 10 to 15 days, which leads to a change in the morphology of the shocked region. The evolution then proceeds uneventfully for tens to hundreds of days. For $\rcsm =0.1 \times 10^{15}$ ($0.3\times 10^{15}$) cm the shock encounters the exponential drop in CBM density at 80 (200) days, while for $\rcsm = 2.0\times 10^{15}$\,cm this will happen only after the end of the simulation. These differences are not important for the hydrodynamics but strongly affect the radiative properties of the outflow, which we discuss in Section~\ref{sec:lc}. 


\subsection{Radiative structure}
\label{sec:flux}

\begin{figure*}
\plottwo{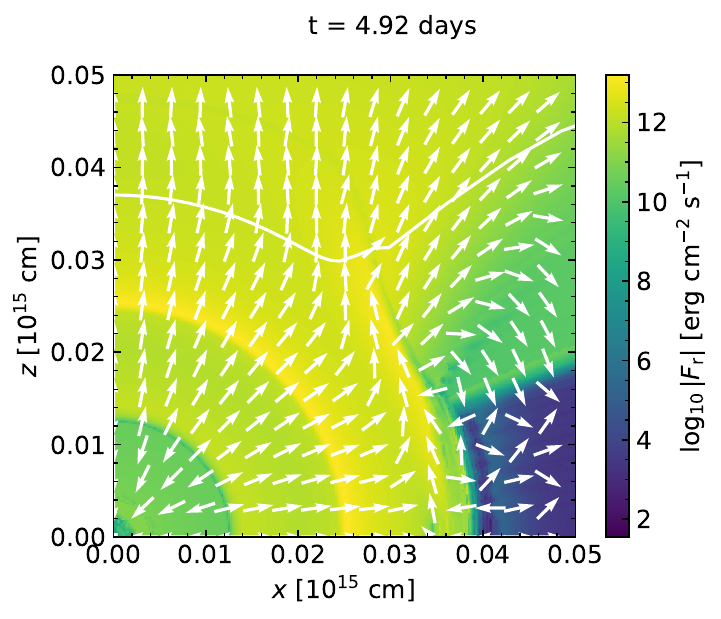}{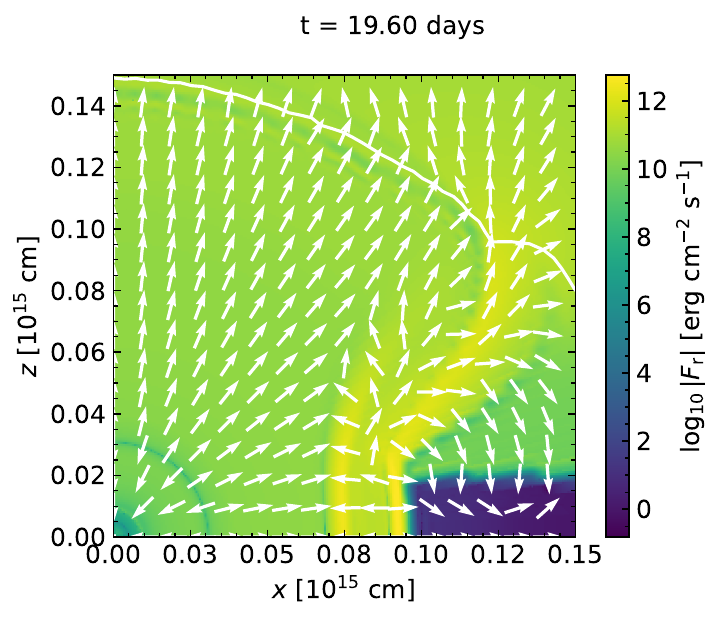}
\plottwo{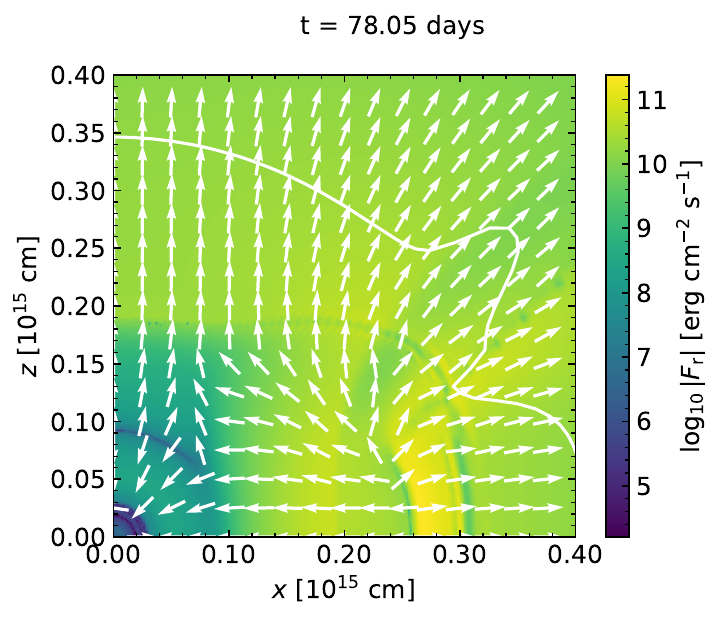}{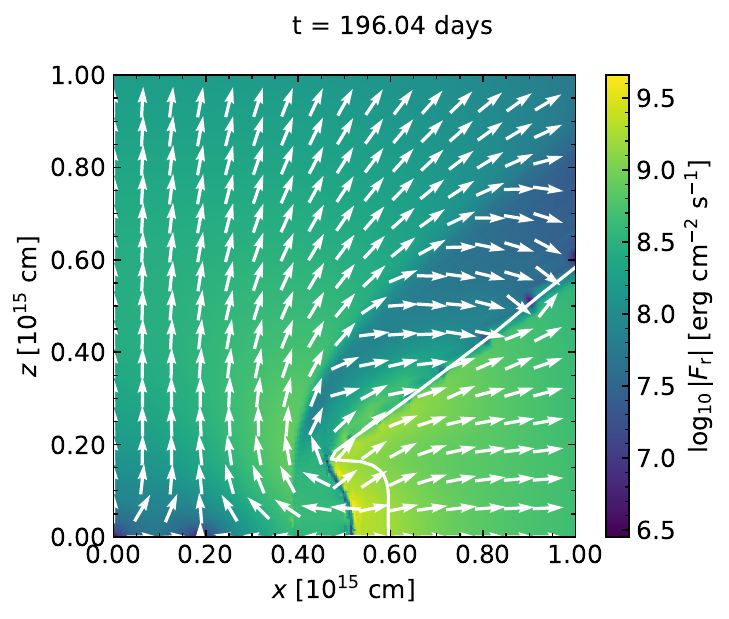}
\caption{Evolution of radiative flux for simulation d0.3hiE with $\tejin=5 \times 10^5$\,K and $\rcsm = 2.0 \times 10^{15}$\,cm. Background color shows the absolute value of the radiative flux while the arrows show its direction. Solid white line shows the position of the $\tau =2/3$ surface measured along radial rays using Rosseland-mean opacity.\label{fig:flux}}
\end{figure*}

In Figure~\ref{fig:flux}, we show the evolution of radiative flux $\mathbf{F}_\text{r}$ (Eq.~[\ref{eq:flux}]) for a representative simulation obtained by postprocessing simulation output of radiation energy density $E_\text{r}$. We show snapshots of the same model and the same four epochs displayed in Figure~\ref{fig:overview}. During the first phase in the first few days (top left panel), we see $\mathbf{F}_\text{r}$ originating in the ejecta--CBM shock propagating laterally and joining radiation emanating radially from the outer layers of the ejecta, $r \gtrsim \rtr$. This flux is responsible for radiatively pushing material away from the equatorial plane, which manifests as negative $v_\theta$ in Figure~\ref{fig:overview}.
Inside the inner ejecta, $r \lesssim \rtr$, $\mathbf{F}_\text{r}$ points inward, which is an artifact of the initial temperature distribution, but this behavior is irrelevant due to the low values of $|\mathbf{F}_\text{r}|$. We also see a bright ring at $2.6\times 10^{13}$\,cm which is present only in high $\tejin$ models and is responsible for the peak of the light curve at $t\approx 5$\,days.
After a few weeks (top right panel), we see that $\mathbf{F}_\text{r}$ points outward in most of the simulation domain indicating that both the shock and diffusion through the ejecta contribute to the emerging luminosity. Radiation of the shock also propagates to and heats up the CBM, although the details of this process are likely affected by the diffusive nature of our radiation transport approximation. After several months (lower left panel), the energy contained in the ejecta decays, but the shock power continues to be high. As a result, radiation flows from the shock into the ejecta, keeping it ionized, and reprocessing the shock power into a relatively isotropic emission. Finally, after most of the ejecta cools down (lower right panel), the shock radiation emerges while being reprocessed by the dusty ejecta.

\subsection{Light curves}
\label{sec:lc}

\begin{figure*}
\plotone{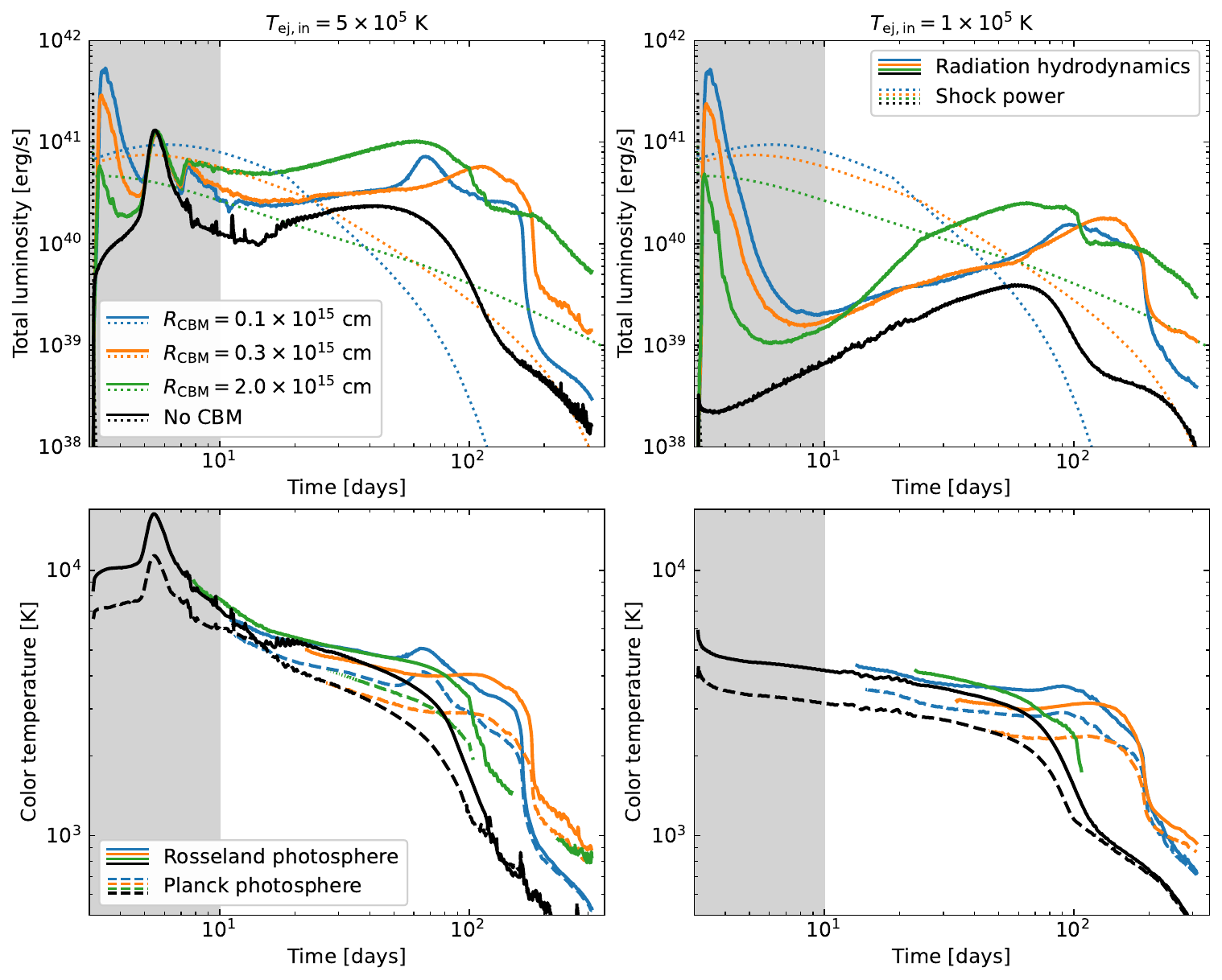}
\caption{Bolometric light curves (upper row) and photospheric temperatures (lower row) of our simulations for the two different initial ejecta temperatures (left and right columns). In the upper row, colored solid lines are for different values of $\rcsm$ while  solid black lines show simulations without CBM. Colored dotted lines indicate shock power for the appropriate ejecta and CBM combination calculated using semi-analytic model for the same ejecta and CBM properties \citep{metzger14,Metzger2017,pejcha22}. In the bottom row, solid lines indicate gas temperatures at Rosseland photosphere and dashed line at Planck photosphere. Photospheric temperatures are shown only for times when the entire photosphere is located on the computational grid. Gray shaded areas indicate regions where the results are significantly affected by the choice of initial conditions.\label{fig:lc}}
\end{figure*}

In the upper row of Figure~\ref{fig:lc}, we show evolution of the bolometric luminosity $\lbol$ in our simulations obtained by integrating the radiative flux in the optically-thin limit, $F_\text{r} = c E_\text{r}$, over the outer edge of the domain. We verified that this calculation gives effectively the same results as using the full Equation~(\ref{eq:flux}) with flux limiters, but is considerably more robust against numerical noise during this post processing of simulation outputs. We see that simulations with CBM show a first peak reaching $\gtrsim 10^{41}$\,\ergss\ at $t$ between 3 to 4\,days. This peak is caused by the initial shock interaction between the ejecta and CBM in optically-thin limit, as we explained in Section~\ref{sec:overall}. This peak is absent in spherically-symmetric runs without CBM. Runs with high $\tejin$ show another peak at $t \approx 5$\,days. This peak occurs even in the run without CBM and is completely absent in all low $\tejin$ runs. The origin of this peak can be explained by realizing that the initial thermal energy content of high $\tejin$ ejecta slightly exceeds the initial kinetic energy and the ejecta has enough thermal energy to hydrodynamically adjust its expansion, leading to a formation of a locally increased radiative flux (bright ring seen at $2.6\times 10^{13}$\,cm in the upper left panel of Fig.~\ref{fig:flux} which is absent in low $\tejin$ models) and a short luminosity peak. Properties of both of these features imply that our results within the first $\sim 10$\,days are significantly affected by the specific choice of our initial conditions.

Runs without CBM show a plateau lasting about 100\,days and peak luminosity in proportion to the initial thermal content of the ejecta. All our CBM runs exhibit a second peak or plateau with durations between $100$ and $200$\,days and luminosities between $10^{40}$ and $10^{41}$\,\ergss. Some of these plateaus exhibit finer features that can be related to the initial conditions. For example, the peak at $t\approx 65$\,days in the  s0.1hiE run can be explained by the shock reaching the exponential cutoff of the CBM and the radiation breaking out of the equatorial region. 

The durations and luminosities of the plateaus are not a simple function of $\rcsm$ and represent the ability of the ejecta--CBM shock to deposit energy in the ejecta. For example, comparison of runs with $\rcsm = 0.1 \times 10^{15}$\,cm and $0.3\times 10^{15}$\,cm shows that even if the shock interaction ends in the middle of plateau, the deposited energy in the ejecta can be sufficiently high to power plateaus of almost the same durations and luminosities. Similarly, the shock power in runs with $\rcsm = 2.0\times 10^{15}$\,cm is insufficient to keep the ejecta ionized, which causes the plateau to end after about $100$\,days. This behavior is consistent with predictions of \citet{matsumoto25} based on semi-analytic models.

The end of the plateau is accompanied by formation of molecular hydrogen and increase of opacities due to dust formation, which also increases again the radius of the photosphere. After the plateau ends, the luminosity evolution is related to the instantaneous shock power, which we calculated using the shocked shell model \citep{metzger14,Metzger2017,pejcha22} with the same ejecta and CBM profiles as used in simulations. The excess luminosity seen in simulations compared to shock power can be explained by cooling of the residual thermal energy of the gas. The details of the late LRN emission are highly dependent on the CBM and wind densities and the possible irradiation by the merger remnant discussed in Section~\ref{sec:remnant}.


Finally, we assess the importance of H and He recombination for the light curves. We performed several runs with a simple $\Gamma$-law EOS and H-only EOS but otherwise identical radiation solvers and initial conditions. The results are shown in Figure~\ref{fig:recomb} in Appendix~\ref{app:tests}. We find that hydrogen recombination in the EOS is able to extend the duration and luminosity of the plateau/second peak, but this is highly dependent on initial conditions. For runs with CBM, hydrogen recombination extends the plateau by only about 20\,days and does not significantly change the luminosity. Without CBM, the effect depends on the ratio of $E_\text{rec}$ to the total thermal energy of the gas. For $\tejin = 5\times 10^5$\,K, the thermal energy is dominated by radiation and recombination has little effect. For $\tejin = 1\times 10^5$\,K, recombination is a substantial contribution to the overall initial energy and the light curves with and without recombination are very different. Differences in plateau luminosities and durations between H-only and H+He EOS are very small.

\subsection{Temperature of the photosphere}
\label{sec:teff}

The key importance of radiative shocks in our simulations motivates search for observational signatures of their presence. One possibility is to study the temperature of radiation emitted. In the lower row of Figure~\ref{fig:lc}, we show the time evolution of the photospheric temperature, which we determined by finding the gas temperature at the photosphere at $\tau =2/3$ along each radial ray and then calculating the angular mean weighted by the radiative flux in a given direction. A more complete view of our models is shown in Appendix~\ref{app:tests}, Figure~\ref{fig:lckappa}. For calculating $\tau$, we can use Rosseland or Planck opacities. Since, typically $\kappa_\text{P} > \kappa_\text{R}$, the Planck photospheres show lower temperatures. Our experiments in Appendix~\ref{app:tests} suggest that Rosseland photospheres more closely match other codes used in literature, but the results are somewhat sensitive to the assumed opacity floor and the true temperatures might be somewhat higher than what we show in Figure~\ref{fig:lc}. We also calculate photospheric temperatures only for epochs when the photosphere is located on our computational grid for all angles.

For $t \lesssim 10$\,days, the photosphere fits onto our grid only for models without the CBM. The photospheric temperatures during this phase are between 6000 and 12000\,K, however, we discussed in Section~\ref{sec:lc} that the behavior at these early times is primarily determined by our initial conditions. Later, until the end of the plateau, photospheric temperatures decrease and level between 3000 and 4000\,K. For model d0.1hiE, we see a small increase in photospheric temperature at $t\approx 60$\,days when the luminosity temporarily increases due to the shock breaking out of the equatorial regions. The end of the plateau is associated with a drop in photospheric temperatures due to dust formation reaching 500 to 1000\,K at the end of our runs.  Photospheric temperatures of shock-powered models are typically slightly higher than for equivalent CBM-free models, but the differences are only several hundred K. Models with lower $\tejin$ consistently show lower photospheric temperaturs. The only informative signature of the presence of shocks is the bump in luminosity and temperature associated with the shock breaking out of the outer edge of the CBM during the plateau phase.

\subsection{Viewing angle dependence}
\label{sec:ray}

\begin{figure*}
    \plottwo{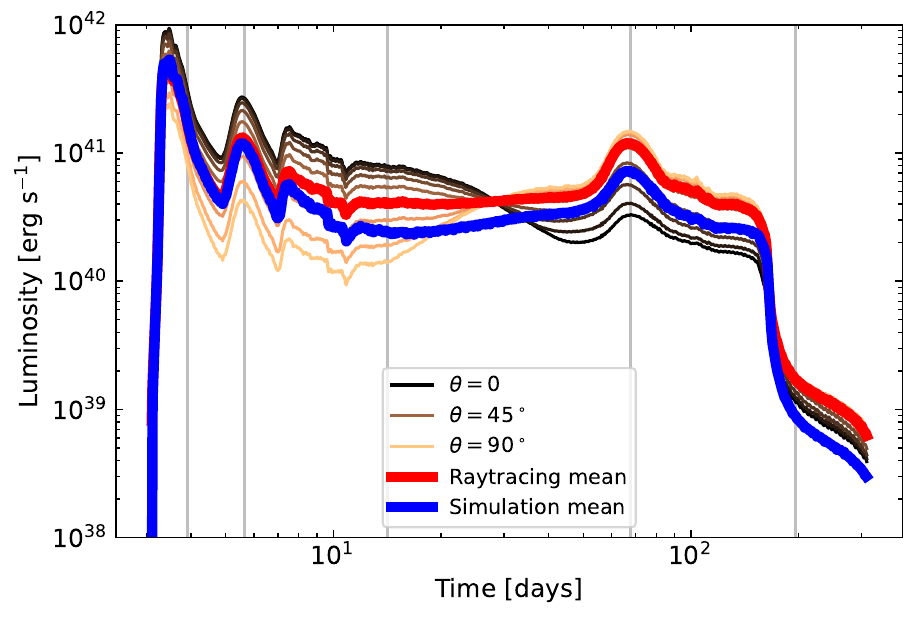}{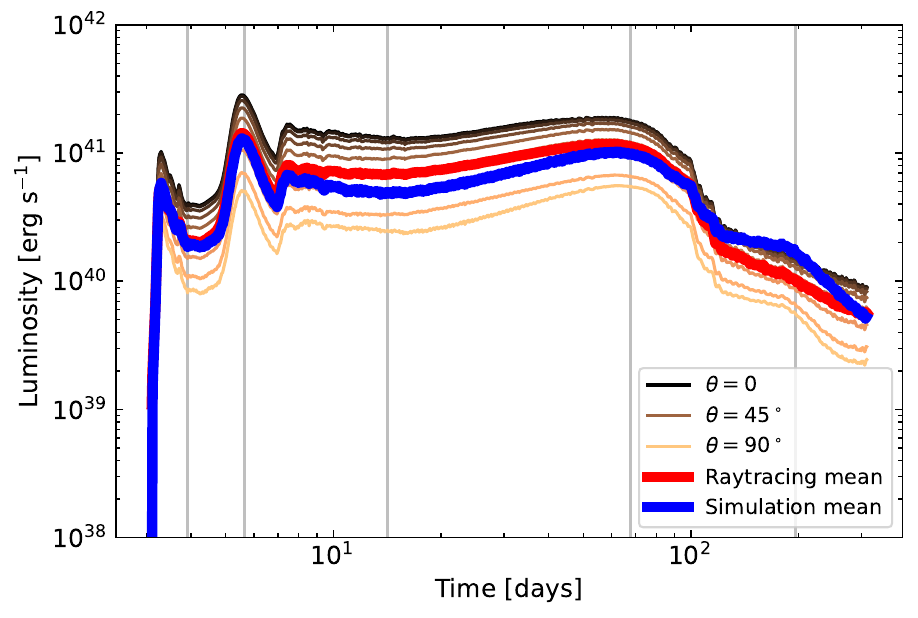}
    \vskip 3mm
    \plottwo{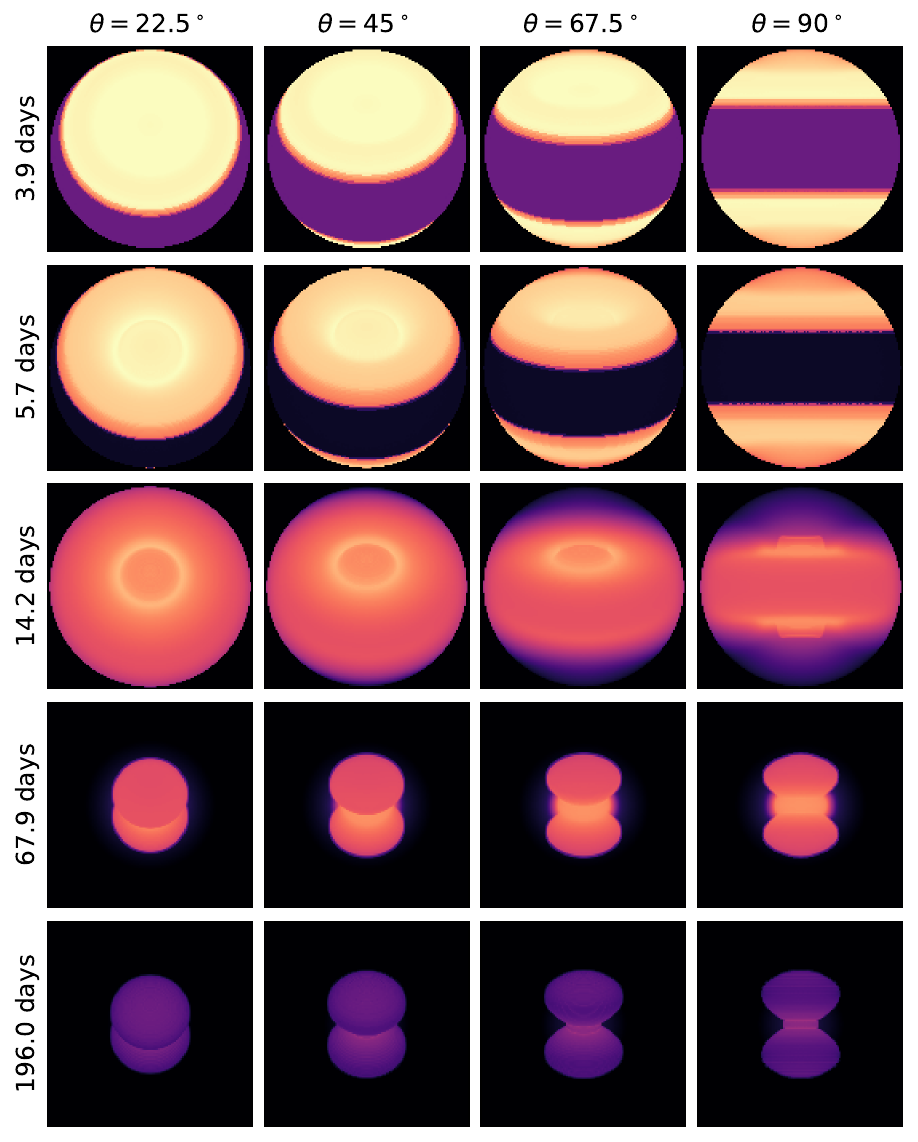}{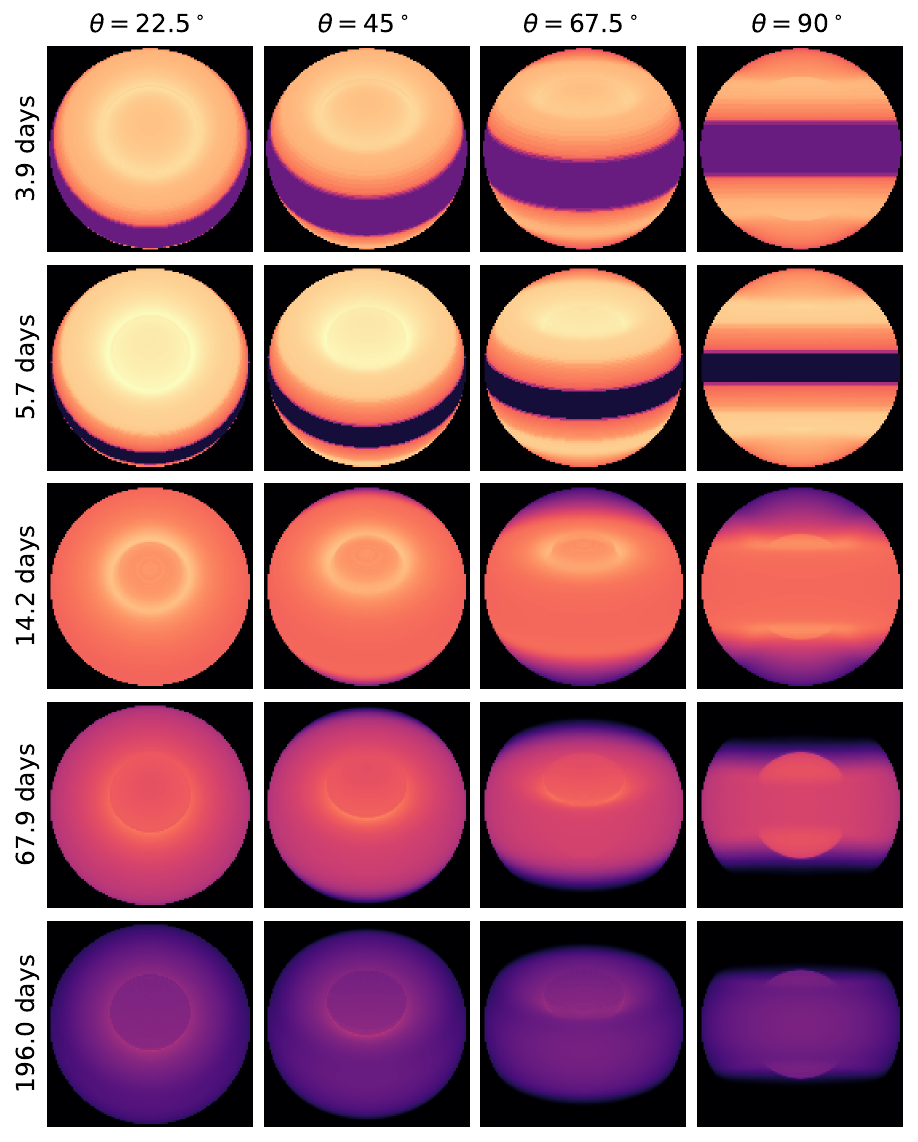}
    \caption{Raytracing of our simulation runs d0.1hiE (left column) and d2.0hiE (right column). Upper row shows evolution of luminosities calculated using raytracing for nine orientation angles $\theta$ (thin lines with colors transitioning from black to orange), angular average of raytracing (thick red line), and the angular average of radiative flux in the simulations (thick blue line, same as in Fig.~\ref{fig:lc}). Lower part shows raytraced images for four selected orientation angles and at five different epochs indicated by thin gray vertical lines in the top row. \label{fig:ray}}
\end{figure*}

In order to assess the viewing angle dependence of our results, we performed raytracing through our simulations.  The technical details of the procedure are presented in Appendix~\ref{app:ray}. In Figure~\ref{fig:ray}, we show results for runs d0.1hiE and d2.0hiE, which differ by the extent of CBM. Animations and results for the remaining runs are available in Appendix~\ref{app:movies}, Figures~\ref{fig:d0.1hiE}--\ref{fig:d2.0loE}. We see that the light curves exhibit considerable viewing angle variation that can reach a factor of 3 from the angular mean. In the first 10 days, the raytraced object looks like a bright sphere with a portion of the equatorial flux blocked by the CBM. The object appears brighter when viewed pole-on. Starting at 15 days, the equatorial ring heats up and becomes more transparent and above it we begin to see a bright ring, which corresponds to reprocessed shock radiation leaving the ejecta. This ring might manifest with double peaked line profiles \citep{gill99}. For model d0.1hiE, the outer edge of the CBM is reached at around 65 days, which leads to a shock breakout visible as a bright equatorial ring. The CBM also stops shaping the ejecta, which continues to expand in the shape of bipolar lobes. As a consequence, the shape of the object changes from oblate to prolate and the object appears appears brighter when viewed edge on, explaining the crossing of lines at $t\approx 30$\,days in the upper left panel of Figure~\ref{fig:ray}. Observing this change of geometry, for example with spectropolarimetry, together with a later correlated bump in luminosity and temperature could serve as and indication of shock breakout out of the outer edge of the CBM. Conversely, model d2.0hiE continues to be surrounded by equatorial CBM, its shape remains oblate, and it remains brighter when viewed pole on.

\subsection{Effect of the radiation from the central remnant}
\label{sec:remnant}

\begin{figure*}
\centering
    \includegraphics[width=\textwidth]{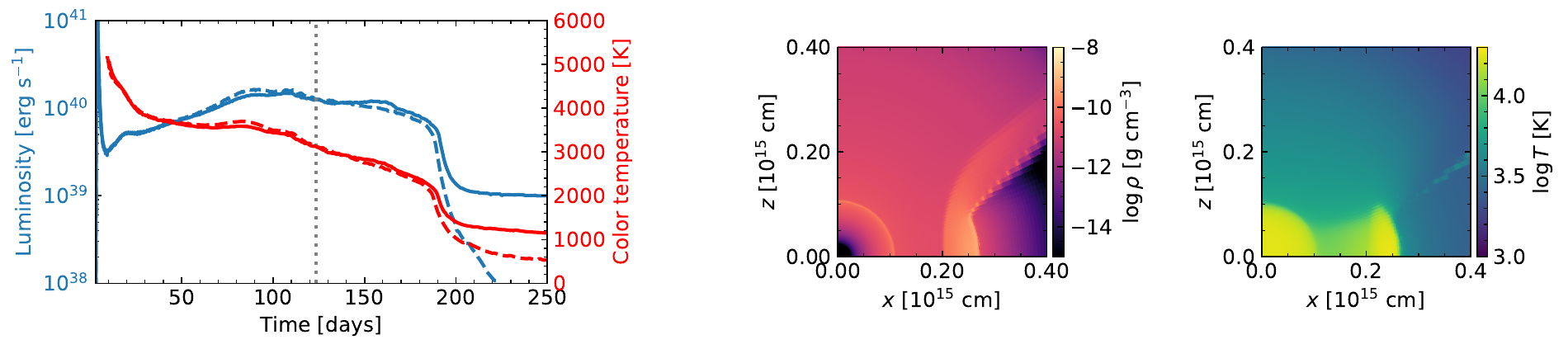}
    \caption{Effect of irradiation from the central source. Left panel shows the luminosities (blue) and photospheric temperatures (red) for the fiducial (dashed, s0.1loE) and irradiated (solid, s0.1loEirr) models. Middle and right panels show the density and gas temperature structure of the irradiated model at one selected epoch indicated by vertical gray dotted line in the left panel.    \label{fig:lc_irr}}
\end{figure*}

Self-consistent simulations of massive stars mergers show that the remnant remains bright for a few years, effectively providing a central source of irradiation with $L_\text{irr} \approx 10^{39}$\,\ergss\ \citep{schneider19}. We performed two runs with $L_\text{irr} = 10^{39}$\,\ergss\ starting either at $t_0$ or later at $t=98$\,days for a model with low $\tejin$ and $\rcsm = 0.1 \times 10^{15}$\,cm to maximize the potential effect of irradiation. The evolution in both cases is similar so we present only the first model (s0.1loEirr).

In Figure~\ref{fig:lc_irr}, we show the evolution of luminosity, photospheric temperature, and a representative zoom-in snapshot of density and gas temperature. We find that about $20$\,days after $L_\text{irr}$ switches on, a hot radiation-dominated bubble develops at the center. This bubble pushes out ejecta gas, leading to very low central densities and the formation of a shock-compressed dense ring visible at $10^{14}$\,cm in Figure~\ref{fig:lc_irr}. Similar behavior albeit at higher energies is seen in simulations of magnetar-powered supernovae \citep[e.g.][]{chen20,suzuki21}. As a result of this evolution, the timestep of the simulation becomes limited by hydrodynamics rather than radiation, leading to almost an order-of-magnitude longer calculation times than in the runs without $L_\text{irr}$. As the ejecta and central bubble expand, the adiabatic losses cause decrease of the bubble temperature despite the continuous addition of energy in the center. When the temperature drops to around $10^4$\,K, opacity drops and the accumulated energy leaves over a timescale of several tens of days. The total amount of accumulated energy, taking into account adiabatic losses, is about $10^{46}$\,erg, which when radiated over 50\,days gives additional luminosity of about $2\times 10^{39}$\,\ergss. This is only a small perturbation to the significantly higher plateau luminosity due to shock interaction and is barely noticeable between 150 and 200\,days in left panel of Figure~\ref{fig:lc_irr}. After the plateau ends, the central irradiation is reprocessed by the dusty ejecta and the luminosity levels at around $L_\text{irr}$ with photospheric temperature of about 1000\,K.

We conclude that central irradiation does not significantly alter properties of shock-powered plateau, but might influence post-plateau evolution or events without significant amounts of CBM. In Section~\ref{sec:disc}, we speculate about several modifications to this scenario, which could lead to different observational consequences.

\subsection{Comparison to observed events}
\label{sec:comparison}

\begin{figure*}
\plotone{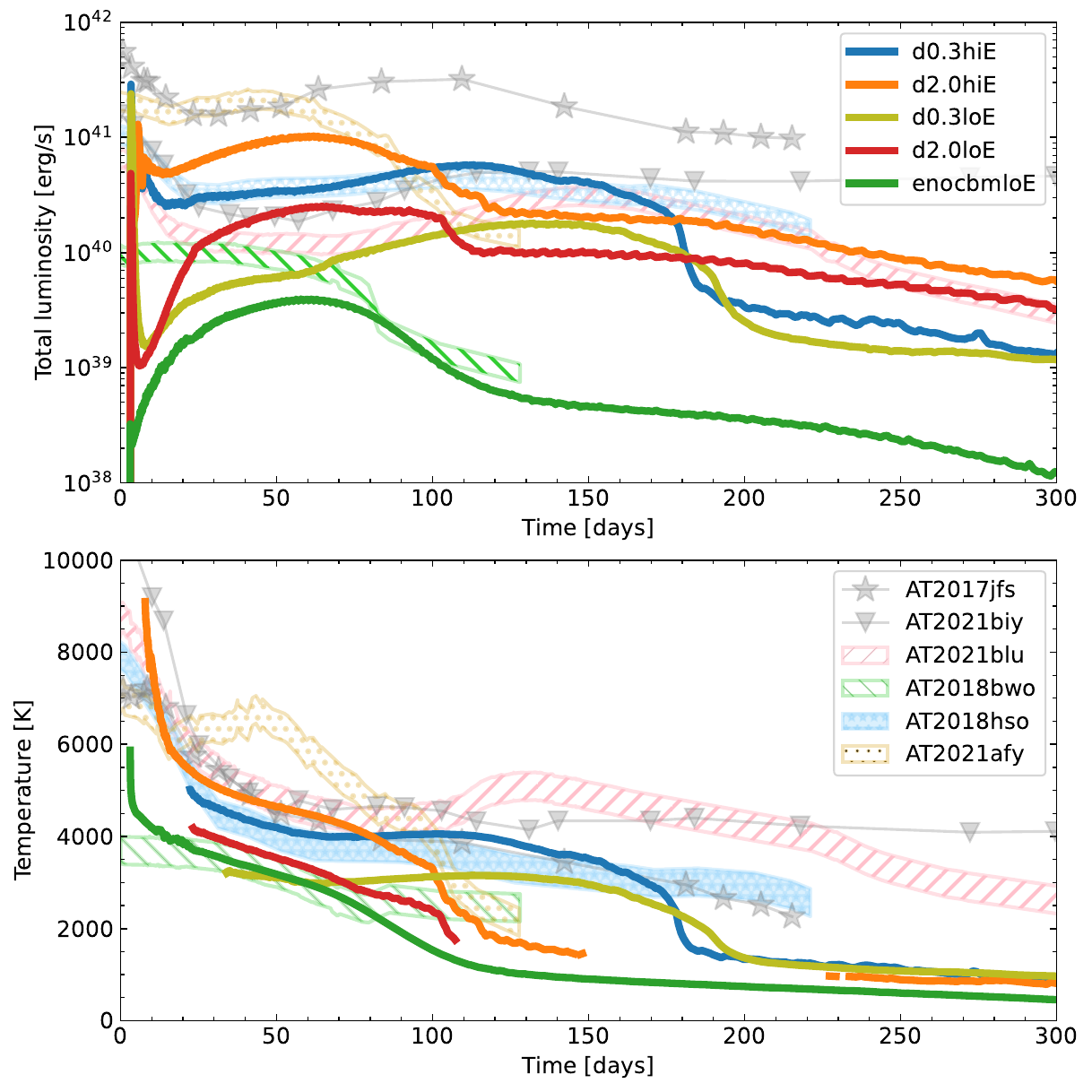}
\caption{Comparison of our simulated luminosities (upper panel) and photospheric temperatures (lower panel) to observed LRNe. Our models are shown with thick solid lines with explanation for the colors given in the upper panel legend. Observed LRNe are AT2017 jfs \citep[gray stars,][]{pastorello19_17jfs}, AT2021biy \citep[gray triangles,][]{cai22}, AT2021blu (pink hatches), AT2018bwo (green hatches), AT2021afy (orange circle hatches) \citep{pastorello23}, and AT2018hso \citep[blue star hatches,][]{cai19_18hso}.    \label{fig:lc_comp}}
\end{figure*}

In Figure~\ref{fig:lc_comp}, we show a comparison of bright extragalactic LRNe to several selected simulations. We emphasize that our models were not constructed to fit any particular event and that the first $10$ to $20$ days of our simulations are significantly affected by our rather arbitrary choice of initial conditions. Consequently, the comparison should be made qualitatively; however, in Figure~\ref{fig:lc_comp} we show models that resemble some of the observed events. At early times, $t \lesssim 50$ days, our CBM models show the typical dip in luminosity separating the first peak from the later plateau, and the peak luminosities of the first peak approximately match the observations, however, all our models show too short durations of the first peak or predict multiple peaks in a short amount of time. It is relevant to remark that in previous sections we discussed that the light curves at $t \lesssim 10$\,days are strongly influence by the initial conditions. Additionally, in Section~\ref{sec:disc}, we present several ways to modify the simulations in order to obtain a better match.

We now discuss individual events in turn, focusing specifically on the luminosity, duration, temperature, and shape of the plateau at $t\gtrsim 50$\,days. AT2021afy (orange circle hatches) finished its plateau at around 10 days, but observations indicate that it might have continued to shine for some time afterward at $\sim 10^{40}$\,\ergss\ with temperatures $\sim 2000$\,K. This behavior is reasonably matched by model d2.0hiE (orange line), where the extended CBM continues to provide shock power after the plateau ends. The luminosities and temperatures of AT2021afy between 30 and 70 days are somewhat higher than in d2.0hiE, but a better match might be obtained by tweaking the CBM density profile and the initial ejecta temperature profile.

AT2018hso (blue star hatches) shows a smooth plateau lasting more than 200 days with nearly constant luminosity of $3\times 10^{40}$\,\ergss\ and slowly declining temperatures. This behavior is well reproduced by the model d0.3hiE (blue line), which finishes the plateau at around 180 days, but extending the plateau might be possible by adjustments to the CBM mass and density profile.

AT2021blu (pink hatches) shows a distinct brightening of the plateau between 100 and 220 days. Similar behavior is shown by model d0.3loE (dark yellow line), except its luminosities before day 70 are somewhat fainter and photospheric temperatures are about 1000 K lower than what is seen in AT2021blu. Perhaps a model with properties intermediate between d0.3loE and d0.3hiE would better match the behavior before 70 days, while changes to CBM mass and density profile could provide longer plateau.

AT2018bwo (green hatches) is the faintest and shortest event displayed, and its plateau properties might be reasonable matched by the CBM-free model enocbmloE if its initial thermal energy was slightly higher. This does not mean that AT2018bwo cannot be explained by shock-powered models; in fact, a more comfortable match in terms of the necessary ejecta mass could probably be achieved by lowering $\mej$ and $\tejin$ of the model d2.0loE (red line).

Finally, we show two additional events that might be harder to explain: AT2017jfs (gray stars) and AT2021biy (gray triangles). AT2017jfs is the brightest event considered with the luminosity of the plateau approaching $3 \times 10^{41}$\,\ergss. Explaining these properties would require an increase in the ejecta mass and velocity and likely also an increase in the CBM mass. Since only $\lesssim 10\%$ of the binary mass is dynamically ejected during a merger, this event would require a high progenitor mass. Unfortunately, progenitor was not detected for AT2017jfs. AT2021biy shows an extremely long plateau that finishes about 300 days after the first peak and a post-plateau bump. The duration of the plateau might stretch any model assuming homologous expansion of ejecta with an additional power source \citep{matsumoto25}. None of our models shows a late-time bump, but we speculate about its possible origin in Section~\ref{sec:disc}.

\section{Summary and Discussions} \label{sec:disc}

In this study, we conducted two-dimensional moving-mesh radiation-hydrodynamic simulations of LRNe, incorporating a hydrogen and helium EOS and realistic opacities. Our initial conditions represent a typical high-mass stellar merger, which ejects $2\,\msun$ of material with a characteristic velocity of $410$\,\kms. This ejecta collides with the preexisting equatorially-concentrated CBM resulting from non-conservative runaway mass leading up to the merger (Fig.~\ref{fig:overview}). We assume CBM mass of about $2.7\,\msun$ concentrated with $\pm 0.1\pi$ of the equator and different radial distributions. 

We find that the bolometric light curves (Fig.~\ref{fig:lc}) begin with a first peak caused by early phases of shock interaction and by cooling of the ejecta with $\lbol \gtrsim 10^{41}$\,\ergss. The initial $\sim$10 days are strongly dependent on the initial conditions of our simulation. The first peak is followed by an extended plateau or a second peak, which lasts up to 200 days and achieves luminosities of $10^{40}$ to $10^{41}$\,\ergss. This second phase is powered by a shock embedded inside the ejecta and we illustrate how the ejecta is re-energized by the radiation originating in the shock (Fig.~\ref{fig:flux}). The photospheric temperatures gradually decrease to around 3000\,K during the plateau and further drop to $\sim 1000$\,K after the plateau ends (Fig.~\ref{fig:lc}). After the end of the plateau, $\lbol$ drops and its further evolution approximately tracks any remaining shock power reprocessed by the dust in the ejecta. The shocks remain embedded during the plateau phase, but their presence might be revealed in situations when the shock reaches the outer edge of the CBM by correlating the change of geometry from oblate to prolate with a consequent bump in luminosity and temperature.

Hydrogen recombination affects our shock-powered results only modestly, but the importance of this effect should increase for lower luminosity events. Helium recombination does not play a noticeable role. Raytracing of the simulations reveals that apparent luminosities can differ from the angular mean by up to a factor of 3 and that the raytraced shape can change from oblate to prolate depending on the extent of the CBM (Fig.~\ref{fig:ray}). Irradiation by the central remnant of the merger creates a radiation dominated low-density bubble, which decays before the end of the plateau, but leaves an imprint on the ejecta density profile and can increase post-plateau luminosities (Fig.~\ref{fig:lc_irr}). 

Our results are broadly consistent with the observed properties of extragalactic LRNe reported in the literature \citep[e.g.,][]{Pastorello2019,Pastorello2021_2019zhd,pastorello23,Blagorodnova2021} (Fig.~\ref{fig:lc_comp}) and with previous semi-analytical models of LRNe \citep{Metzger2017,matsumoto25}. In particular, we are able to simultaneously explain the high luminosities and very long durations of the plateau (Fig.~\ref{fig:lc_comp}), which otherwise requires unrealistically high ejecta masses in spherical models with freely expanding ejecta \citep{Matsumoto2022}. 

There are still features that are unexplained and that motivate future improvements and extensions of our model. First, the duration of the first peak in our simulations is considerably shorter than that observed. Since the properties of the first peak strongly depend on the initial conditions, this discrepancy motivates exploration of different physical effects and more complicated ejecta profiles than the simple broken power law used here. For example, surrounding the region of initial shock interaction with optically-thick gas, such as might potentially arise from immediate pre-merger interactions or due to circumbinary disk winds, could lead to reprocessing of the radiation and extension of the duration of the first peak. Furthermore, a realistic ejecta structure could be obtained from self-consistent three-dimensional simulations of the merger process \citep[e.g.,][]{schneider19,hirai21} and mapped to initial conditions in our code.

Second, the plateau luminosities of some events such as NGC 4490-2011OT1 \citep{Smith2016,Pastorello2019} or AT2021afy \citep{pastorello23} reach $3 \times 10^{41}$\,\ergss, which is by a factor of few higher than what is seen in our models. However, our models were calculated for a limited set of parameters, and considering a wider range $\mej$, $E_\text{ej,kin}$, $\mcsm$, and $\rcsm$ could explain these brightest events. 

Third, our prescription for the radial and vertical structure of the CBM does not specifically represent any physical processes responsible for the pre-merger mass loss from the binary. Further exploration of different analytic prescriptions and more physically motivated profiles representing L2 outflows, circumbinary disks, disk winds, and jets should be performed in an effort to identify unique signatures connecting the observed diversity LRN light curves to the underlying physical processes. This is an important missing link that could together with the pre-merger build up of the CBM provide unique tests of binary evolution models by quantifying the amounts and timing of mass and angular momentum loss from the binary.

Fourth, although irradiation by the central source with $L_\text{irr} \approx 10^{39}$\,\ergss\ does not lead to nontrivial outcomes in the light curve, it might be worth considering more extreme situations that might occur, for example, when the binary interaction outcome involves a compact object accreting and radiating at super-Eddington rates. In such situations, higher $L_\text{irr}$ could compress and accelerate the inner regions of the ejecta more strongly, and the resulting shell could collide with the already dense CBM interaction shell, potentially causing a post-plateau bump in the light curve. The energetics of the bump seen in AT2021biy, $2\times 10^{40}$\,\ergss\ over 50 days for a total of $\sim 10^{47}$\,erg radiated \citep{cai22}, require collision of shells with masses $\gtrsim \msun$ and relative velocities $\gtrsim 100$\,\kms. These requirements, combined with the exceptionally long plateau, make AT2021biy an extreme object worthy of further investigation.


Finally, further developments of post-processing of our simulation would be useful especially for diagnosing the signatures of the shock. 
Photoionization modeling of the shock surroundings would inform on shock emission lines emerging from the ejecta and would help to interpret existing and planned late-time JWST observations of LRNe \citep{regiutti25}.  Our simulation results could also be used as a starting point for theoretical predictions of spectropolarimetry already performed for several LRNe \citep{desidera04,cai22}.

\begin{acknowledgments}
We thank Andrea Pastorello for providing us with bolometric light curves of LRNe, Kengo Tomida for sharing tabulates EOS and opacities, and the referee for detailed comments and suggestions, which significantly improved and expanded this work. We thank B.~Metzger and T.~Kami\'{n}ski for comments on the draft. The research of AK, DC, and OP has been supported by the Horizon 2020 ERC Starting Grant ‘Cat-In-hAT’ (grant agreement no. 803158). 
The research of DC has been funded by the Alexander von Humboldt Foundation. 
AK acknowledges DAAD for funding a visit in Hamburg. OP acknowledges support from the Charles University Research Center Grant No. UNCE/24/SCI/016.
\end{acknowledgments}
\software{SciPy \citep{2020SciPy-NMeth}, Matplotlib \citep{Hunter:2007}, GNU Scientific Library \citep{gsl}
          }


\restartappendixnumbering 

\appendix
\section{Numerical methods}
\label{app:code}

\label{sec:basiceq} 
RJET solves the radiation hydrodynamics equations under the flux-limited diffusion approximation in the mixed frame formulation to ensure high precision energy conservation \citep{Krumholz2007}. Assuming Planckian emission and working with frequency-integrated quantities (gray approximation), keeping terms up to order $\mathcal{O}(u/c)$ \citep{Zhang2011}, the equations are
\begin{eqnarray}
\frac{\partial \rho}{\partial t}+\nabla \cdot(\rho \mathbf{u})&=&0 \label{eq:app_first}\\
\frac{\partial(\rho \mathbf{u})}{\partial t}+\nabla \cdot(\rho \mathbf{u u})+\nabla p+\lambda \nabla E_{\mathrm{r}}&=&\mathbf{0} \label{eq:app_mom} \\
\frac{\partial(\rho E)}{\partial t}+\nabla \cdot(\rho E \mathbf{u}+p \mathbf{u})+\lambda \mathbf{u} \cdot \nabla E_{\mathrm{r}} &=& -c \kappa_{\mathrm{P}}\left(a_{\mathrm{r}} T^4-E_{\mathrm{r}}^{(0)}\right) \\
\frac{\partial E_{\mathrm{r}}}{\partial t}+\nabla \cdot\left(\frac{3-f}{2} E_{\mathrm{r}} \mathbf{u}\right)-\lambda \mathbf{u} \cdot \nabla E_{\mathrm{r}}&=&c \kappa_{\mathrm{P}}\left(a_{\mathrm{r}} T^4-E_{\mathrm{r}}^{(0)}\right) + \nabla \cdot\left(\frac{c \lambda}{\chi_{\mathrm{R}}} \nabla E_{\mathrm{r}}\right),
\label{eq:FLDeq}
\end{eqnarray}
where $\rho, \bf{u}, \mathrm{E_r}, \mathrm{E}$ are density, velocity, radiation energy density and total energy per unit mass ($E=\frac{1}{2} u^2 + \mathrm{\epsilon}$), $\kappa_\text{P}$ is the Planck-mean absorption coefficient\footnote{In Equations~(\ref{eq:app_first}--\ref{eq:app_last}), we use $\kappa_\text{P}$ to refer to the Planck absorption coefficient with units of cm$^{-1}$. In the remainder of the paper, we use $\kappa_\text{P}$ to refer to Planck mean opacity with units of cm$^2$\,g$^{-1}$. This difference is used to maintain consistency both with the literature on radiative transport methods including our previous work \citep{Zhang2011,Calderon2021} and the more general astronomical literature.}, $\chi_\text{R} = \kappa_\text{R} \rho$ is the Rosseland-mean coefficient that includes the contribution of both absorption and scattering processes. The flux limiter $\lambda$ \citep{Levermore1981}, Eddington factor $f$, and the relation between co-moving and lab frame radiation energy density \citep{Zhang2011} are
\begin{equation}
\lambda(R) = \frac{2+R}{6+3R+R^2}, \quad R = \frac{\nabla E_{\mathrm{r}}^{(0)}}{\chi_{\mathrm{R}} E_{\mathrm{r}}^{(0)}}, \quad f = \lambda + \lambda ^2 R^2, \quad E_{\mathrm{r}}^{(0)}=E_{\mathrm{r}}+2 \frac{\lambda}{\chi_{\mathrm{R}}} \frac{\mathbf{u}}{c} \cdot \nabla E_{\mathrm{r}}+\mathcal{O}\left(\frac{u^2}{c^2}\right).
\end{equation}
Under the flux-limited diffusion approximation \citep{alme1973}, the radiative flux is defined as
\begin{equation}
\mathbf{F}_\text{r}^{(0)} = -\frac{c\lambda}{\chi_\text{R}}\nabla E_\text{r}^{(0)}.
\label{eq:flux}
\end{equation}

In the rest of this Appendix, we briefly sketch out how Equations~(\ref{eq:FLDeq}--\ref{eq:flux}) are solved in RJET. In Section~\ref{app:implicit}, we describe how the system is split in two parts, a hyperbolic subsystem that can be solved explicitly via Godunov's method assuming the relation $(3-f)/2 = \lambda + 1$ holds \citep[see][]{Zhang2011} and a parabolic part (the diffusion, and source and sink terms) that is solved implicitly. In Section~\ref{app:refinement}, we describe the algorithm for timestepping and mesh refinement. In Section~\ref{app:eos}, we describe our EOS. In Section~\ref{app:opacity}, we present our choice of opacities. In Section~\ref{app:tests}, we validate our code on Type IIP supernova light curve and discuss results of our resolution study. Finally, in Section~\ref{app:ray}, we present the implementation of the raytracing through our simulations.



\subsection{(Non-linear) Implicit radiation solver}
\label{app:implicit}
Instead of the linear implicit solver previously used in \citet{Calderon2021,Caldron2024}, we implemented the approach developed by \cite{Howell2003} as well as used in other codes such as CASTRO \citep{Zhang2011}. 
The equations of the implicit step involve the internal energy density $e$ and the radiation energy density $E_{\mathrm{r}}$,
\begin{eqnarray}
\frac{\partial e}{\partial t} & =&-c \kappa_{\mathrm{P}}\left(a_{\mathrm{r}} T^4-E_{\mathrm{r}}\right)+2 \lambda \frac{\kappa_{\mathrm{P}}}{\chi_{\mathrm{R}}} \mathbf{u} \cdot \nabla E_{\mathrm{r}}, \\
\frac{\partial E_{\mathrm{r}}}{\partial t} & =&+c \kappa_{\mathrm{P}}\left(a_{\mathrm{r}} T^4-E_{\mathrm{r}}\right)-2 \lambda \frac{\kappa_{\mathrm{P}}}{\chi_{\mathrm{R}}} \mathbf{u} \cdot \nabla E_{\mathrm{r}}+\nabla \cdot\left(\frac{c \lambda}{\chi_{\mathrm{R}}} \nabla E_{\mathrm{r}}\right),
\end{eqnarray}
where $T$ is the temperature of the gas and $c$ is the speed of light. Discretising the equations and replacing $e=\rho \epsilon$, where $\epsilon$ is the specific internal energy, we obtain
\begin{eqnarray}
\frac{\rho \epsilon^{n+1}-\rho \epsilon^{-}}{\Delta t} & =&-c \kappa_{\mathrm{P}}^{n+1}\left[a_{\mathrm{r}}\left(T^{n+1}\right)^4-E_{\mathrm{r}}^{n+1}\right]+2 \lambda^{n+1} \frac{\kappa_{\mathrm{P}}^{n+1}}{\chi_{\mathrm{R}}^{n+1}} \mathbf{u} \cdot \nabla E_{\mathrm{r}}^{n+1} \\
\frac{E_{\mathrm{r}}^{n+1}-E_{\mathrm{r}}^{-}}{\Delta t}&= & +c \kappa_{\mathrm{P}}^{n+1}\left[a_{\mathrm{r}}\left(T^{n+1}\right)^4-E_{\mathrm{r}}^{n+1}\right]-2 \lambda^{n+1} \frac{\kappa_{\mathrm{P}}^{n+1}}{\chi_{\mathrm{R}}^{n+1}} \mathbf{u} \cdot \nabla E_{\mathrm{r}}^{n+1}+\nabla \cdot\left(\frac{c \lambda^{n+1}}{\chi_{\mathrm{R}}^{n+1}} \nabla E_{\mathrm{r}}^{n+1}\right) .
\end{eqnarray}
In order to solve the system using Newton's method we define
\begin{eqnarray}
F_e &= & \ \rho \epsilon^{n+1}-\rho \epsilon^{-} -\Delta t \left\{ - c \kappa_{\mathrm{P}}^{n+1}\left(a_{\mathrm{r}}\left(T^{n+1}\right)^4-E_{\mathrm{r}}^{n+1}\right) + q^{n+1} \mathbf{u} \cdot \nabla E_{\mathrm{r}}^{n+1} \right\}\\
F_{\mathrm{r}} &= & \ E_{\mathrm{r}}^{n+1} - E_{\mathrm{r}}^{-} - \Delta t \left\{ c \kappa_{\mathrm{P}}^{n+1} \left[ a_{\mathrm{r}} \left( T^{n+1} \right) ^4-E_{\mathrm{r}}^{n+1} \right]-q^{n+1} \mathbf{u} \cdot \nabla E_{\mathrm{r}}^{n+1}\right\} - \Delta t \ \nabla \cdot \left( d^{n+1} \nabla E_{\mathrm{r}}^{n+1} \right) 
\end{eqnarray}
where $q^{n+1}=2 \lambda^{n+1} \kappa_{\mathrm{P}}^{n+1} / \chi_{\mathrm{R}}^{n+1}$ and $d^{n+1}=c \lambda^{n+1} / \chi_{\mathrm{R}}^{n+1}$.
Then, we construct the following linear system that needs to be solved iteratively using the superscript $k$,
\begin{equation}
\begin{bmatrix}
\frac{\partial F_{e_k}}{\partial e} & \frac{\partial F_{e_k}}{\partial E_{\mathrm{r}}} \\
\frac{\partial F_{\mathrm{r}}}{\partial e} & \frac{\partial F_{\mathrm{r}}}{\partial E_{\mathrm{r}}}
\end{bmatrix}
\begin{bmatrix}
\delta e^{(k+1)} \\
\delta E_{\mathrm{r}}^{(k+1)}
\end{bmatrix}
=
\begin{bmatrix}
-F_e^{(k)} \\
-F_{\mathrm{r}}^{(k)}
\end{bmatrix},
\end{equation}
where $\delta e^{(k+1)}=e^{n+1,(k+1)}-e^{n+1,(k)}$ and $\delta E_{\mathrm{r}}^{(k+1)}=E_{\mathrm{r}}^{n+1,(k+1)}-E_{\mathrm{r}}^{n+1,(k)}$. Making use of the Schur complement to eliminate the temperature dependence, and dropping the $n+1$ superscript,
\begin{equation}
\left[{\frac{\partial F_{\mathrm{r}}}{\partial E_{\mathrm{r}}}}^{(k)}+\eta{\frac{\partial F_e}{\partial E_{\mathrm{r}}}}^{(k)}\right]\left[E_{\mathrm{r}}^{(k+1)}-E_{\mathrm{r}}^{(k)}\right]=-F_{\mathrm{r}}^{(k)}-\eta F_{\mathrm{e}}^{(k)},
\end{equation}
where
\begin{equation}
\eta  =-\frac{\partial F_{\mathrm{r}}}{\partial e}\left(\frac{\partial F_e}{\partial e}\right)^{-1}  =-\frac{\partial F_{\mathrm{r}}}{\partial T}\left(\frac{\partial e}{\partial T}\right)^{-1}\left[\frac{\partial F_e}{\partial T}\left(\frac{\partial e}{\partial T}\right)^{-1}\right]^{-1}  \approx 1-\frac{\frac{\partial e}{\partial T}}{\frac{\partial e}{\partial T}+c \Delta t \frac{\partial}{\partial T}\left[\kappa_{\mathrm{P}}\left(a_{\mathrm{r}} T^4-E_{\mathrm{r}}\right)\right]}.
\end{equation}
It is important to remark that density $\rho$ remains as a constant during the implicit step.
After some algebraic manipulation, we obtain an equation that needs to be solved for $E_{\mathrm{r}}^{(k+1)}$ in every Newton iteration,
\begin{equation}
\begin{split}
\left[1+\Delta t(1-\eta) c \kappa_{\mathrm{P}}\right] E_{\mathrm{r}}^{(k+1)} -\Delta t \nabla \cdot\left(d \nabla E_{\mathrm{r}}^{(k+1)}\right) +\Delta t(1-\eta) q \mathbf{u} \cdot \nabla E_{\mathrm{r}}^{(k+1)} =\\
=\Delta t(1-\eta) c \kappa_{\mathrm{P}} a_{\mathrm{r}}\left(T^{(k)}\right)^4+E_{\mathrm{r}}^{-}-\eta\left(\rho \epsilon^{(k)}-\rho \epsilon^{-}\right).
\end{split}
\end{equation}
Integrating over the volume of an arbitrary cell of volume $V$ and using Gauss' theorem,
\begin{equation}
\begin{split}
\left[\frac{1}{\Delta t}+(1-\eta)\left(c \kappa_{\mathrm{P}}-q \nabla \cdot \mathbf{u}\right)\right] E_{\mathrm{r}}^{(k+1)} -\frac{(1-\eta) q}{V} \sum_f^{\text {faces }} E_{\mathrm{r}_f}^{(k+1)} \mathbf{u}_f \cdot \mathbf{A}_f -\frac{1}{V} \sum_f^{\text {faces }} d_f \nabla E_{\mathrm{r}_f}^{(k+1)} \cdot \mathbf{A}_f = \\
= (1-\eta) c \kappa_{\mathrm{P}} a_{\mathrm{r}}\left(T^{(k)}\right)^4+\frac{1}{\Delta t}\left[E_{\mathrm{r}}^{-}-\eta\left(\rho \epsilon^{(k)}-\rho \epsilon^{-}\right)\right],
\end{split}
\label{LinearSystem}
\end{equation}
where $\mathbf{A}_f$ represents an arbitrary face of a given cell and the subscript $f$ runs over each face. This equation represents a system of linear equations that needs to be solved for $E_{\mathrm{r}}^{(k+1)}$ in every Newton iteration. Afterwards, the gas internal energy density needs to be updated accordingly through
\begin{equation}
\rho \epsilon^{(k+1)} = \ \eta \rho \epsilon^{(k)}+(1-\eta) \rho \epsilon^{-} - \Delta t(1-\eta) c \kappa_{\mathrm{P}}\left[a_{\mathrm{r}}\left(T^{(k)}\right)^4-E_{\mathrm{r}}^{(k+1)}\right] + \Delta t \  q \mathbf{u} \cdot \nabla E_{\mathrm{r}}^{(k+1)}.  \label{eq:app_last}
\end{equation}
Finally, $T^{(k+1)}$ can be obtained from the EOS. The remaining quantities ($\kappa_{\mathrm{P}}, \chi_{\mathrm{R}}, \eta, d$, and $q$) are also updated after every Newton iteration, because they depend on $T$. In the same way, $\lambda$ is recalculated using $E_{\mathrm{r}}^{(k+1)}$. The iterations are stopped once the maximum of $\mathcal{E} = |E_{\mathrm{r}}^{(k+1)}-E_{\mathrm{r}}^{(k)}| / E_{\mathrm{r}}^{(k+1)}$ across the entire computational domain gets smaller than a predefined tolerance, which we set to $1 \times 10^{-4}$ in most of our runs. We do not need to decrease tolerance of linear solver with every Newton iteration as we solve for $E_{\mathrm{r}}^{(k+1)}$ and not the change \citep{Dembo1982}.  

The system of linear equations (\ref{LinearSystem}) is solved using the Transpose-Free Quasi-Minimal Residual Method (TFQMR) iterative solver from the PETSc library \citep{petsc-web-page,petsc-user-ref,petsc-efficient}. We performed a test showing the results are similar as using the stabilized Bi-Conjugate gradient (BiCGSTAB) method \citep{vandervorst1992}. In both cases, we also made use of the boomerAMG preconditioner \citep{henson2002} from the library \textit{hypre} \citep{falgout2006} as this was shown to aid in convergence and stability.

\subsection{Timestep and Refinement}
\label{app:refinement}

Given a value of $\Delta t$, the non-linear solver attempts to find the solution. If this is not successful, typically because $\mathcal{E}$ does not monotonically decrease within a specific number of iterations, the solver will restart the implicit step but decreasing $\Delta t$ by a factor of two, forming an inner iteration subcycle similar to the one in \citet{Zhang2012} with a sub-timestep $\Delta t'$. If the solver still does not converge it will decrease $\Delta t'$ again until it is small enough and manages to converge. Then, the solver repeats the substeps until $\Delta t$ is reached with multiple $\Delta t'$. If convergence improves during these substeps, measured by the number of required iterations, $\Delta t'$ can increase to reach $\Delta t$ quicker. We further implement a timestep selection algorithm using a low-pass filter informed from digital control theory \citep{Soderlind2006} and similar to the one implemented in MESA \citep{Paxton2010}. This algorithm includes results of the previous timestep and relative change of quantities.

We further implemented radial adaptive-mesh refinement, where a cell is split if the relative temperature (or $E_\text{r}$) difference with its neighbors is greater than a predefined tolerance with a default value of 0.03 \citep[similar to][]{Chen24}.
The existing refinement in JET limits the ratio of the radial to angular size of cells to be between preset values, which we set to be between 1/2 and 10 for the default initial number of radial zone $N_r = 512$. The combination of the two refinement criteria allows cells in outer regions to grow longer, which aids with the convergence of Newton iterations \citep[see Appendix A in][]{Tetsu2016}, while retaining resolution where gradients are large.

\subsection{General EOS in RJET}
\label{app:eos}

\begin{figure}
    \centering
    \includegraphics[width=0.7\textwidth]{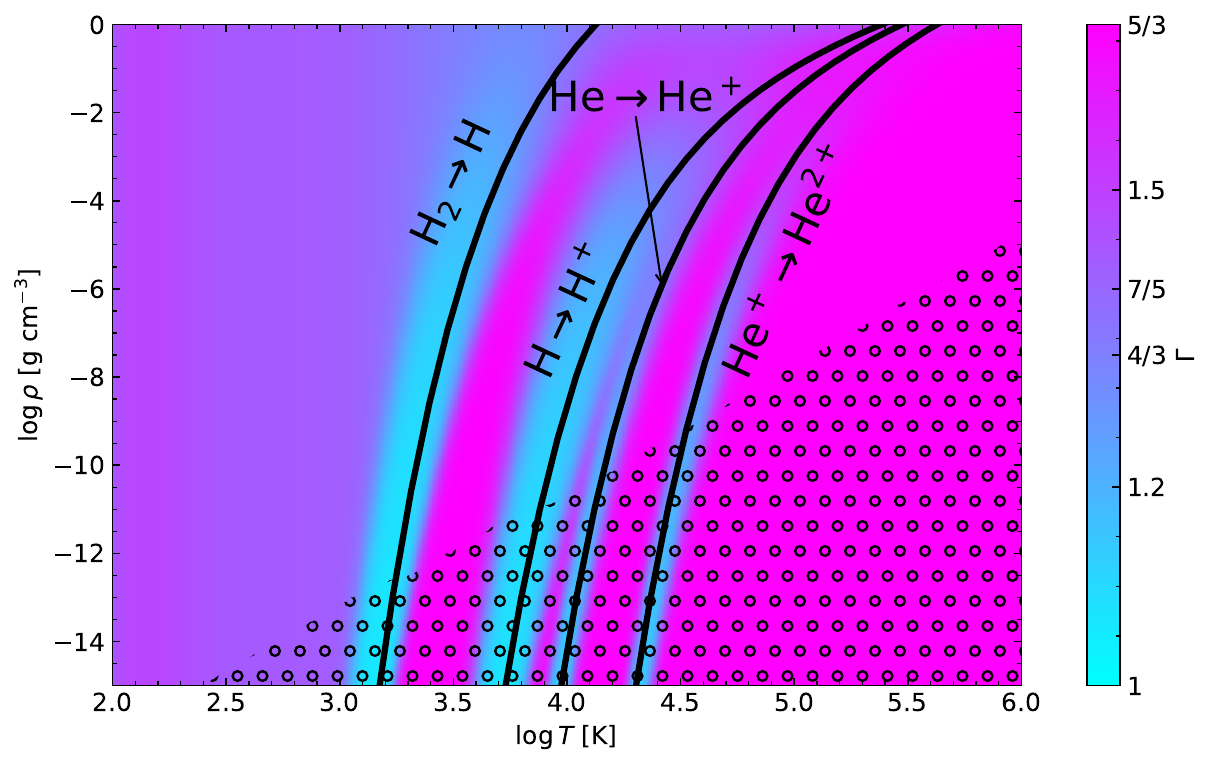}
    \caption{Adiabatic index $\Gamma$ of our tabulated equation of state \citep{tomida13,tomida15} as a function of $\rho$ and $T$. Black solid lines show empirical Saha-like functions that we used to determine the ionization state of gas and black circles indicate region where $a_\text{r}T^4/3$ exceeds gas pressure (see Figs.~\ref{fig:overview} and \ref{fig:d0.1hiE}--\ref{fig:d2.0loE}). \label{fig:eos}}
\end{figure}

Our simulations use three different EOS: a simple $\Gamma$ law, analytic H recombination/ionization, and a general tabulated EOS with H and He physics. The latter two options require special treatment compared to the original version of JET, which we describe below. In general, both H analytic and H+He tabulated EOS assume that all of the recombination energy is fully thermalized in the gas which is then coupled to radiation through the Planck-mean opacity. Since recombination produces photons, it would be more appropriate to deposit the recombination energy in radiation, however, this would require multi-group radiation transport and would introduce other short timescales into the problem.

RJET's general equation of state extension was implemented in a similar manner as was done in Athena++ \citep{Coleman2020}. We use a primitive variable solver where we specify $p(\rho, e)$ and $e(\rho, p)$ analytically. Then, the temperature is found when a conversion from conservatives to primitives is needed. 

\subsubsection{Analytic hydrogen EOS}
We use an analytical equation of state that includes energy from recombination of hydrogen, 
\begin{equation}
e(\rho, T) = \frac{3}{2}(1+\xion) \frac{\rho}{m_p} k_{\mathrm{B}}\ T + \frac{\rho}{m_p} \xion \varepsilon_i, \quad P(\rho, T) = (1+\xion) \frac{\rho}{m_p} k_{\mathrm{B}} T
\end{equation}
where $\xion$ is the hydrogen ionization fraction obtained by solving the Saha equation, and $\varepsilon_i = 13.6$\,eV is the energy of ionization.

\subsubsection{Tabulated hydrogen and helium EOS}

We use tabulated EOS calculated by \citet{tomida13,tomida15}, which we also used in our previous work on binary outflows \citep{pejcha16_cool,pejcha16_bound,Pejcha2017}. In Figure~\ref{fig:eos}, we summarize the basic features of this EOS using the adiabatic index $\Gamma$. This EOS assumes equilibrium abundances of ionized, neutral, and molecular hydrogen ($X=0.7$) and neutral, singly, and doubly ionized helium ($Y=0.3$). The treatment of molecular hydrogen includes rotational and vibrational degrees of freedom assuming thermal equilibrium of ortho- and para-H$_2$, however, these low-temperature features are inconsequential for the treatment of the bright LRNe discussed here, but might become important when studying later phases of their evolution or less-energetic infrared transients. The EOS is tabulated on a densely spaced logarithmic grid that spans $-22 \le \log \rho \le 1.1$ and $0.2 \le \log T \le 8$. We use bi-log-linear interpolation and Brent's method root finder to perform the necessary EOS operations. The range of our table is fully sufficient for our simulations and attempts to access values out of this range typically indicate problems with solution convergence or numerical noise.


\subsection{Opacity}
\label{app:opacity}

\begin{figure}
\centering
    \includegraphics[width=\textwidth]{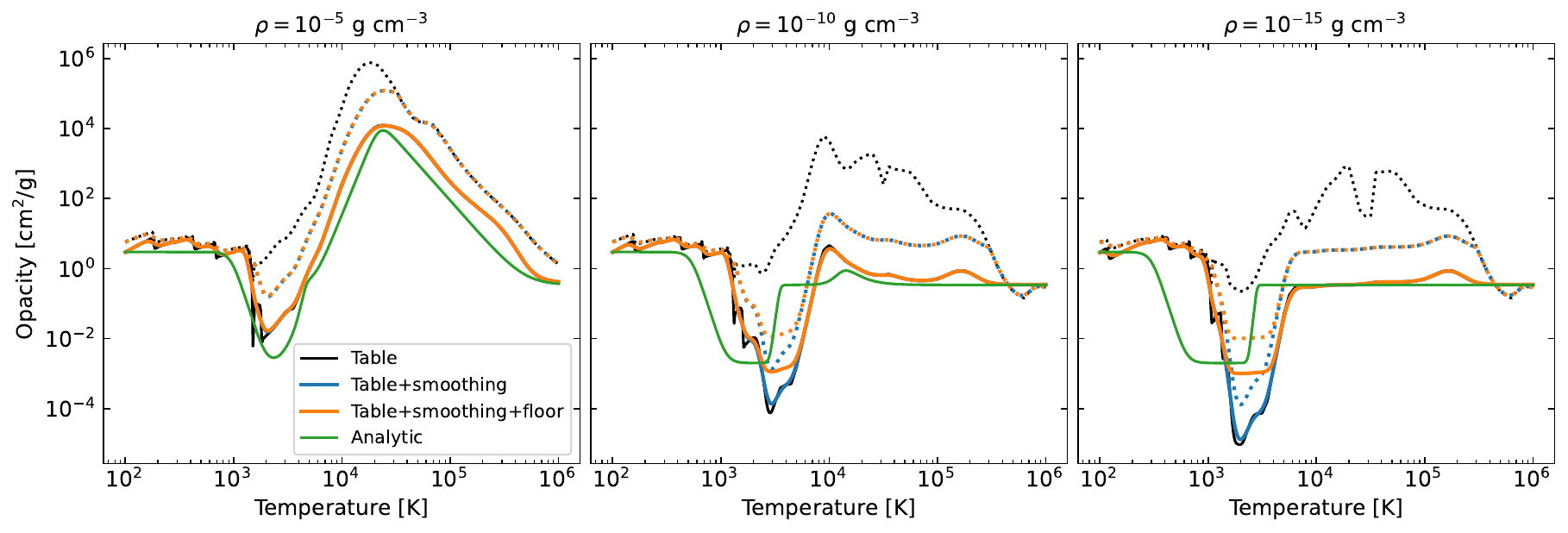}
    \caption{Comparison of opacity prescriptions used in this work. Each panel shows temperature dependence of the opacity for a different density. Solid lines denote Rosseland means and dotted lines Planck means. Black lines shows original tables from \citet{tomida13}, blue lines apply smoothing, orange lines further  apply Rosseland mean floor of $10^{-3}$\,cm$^2$\,g$^{-1}$, and green lines show analytic prescription from Eq.~(\ref{eq:opacity}). \label{fig:opacity}}
\end{figure}


The choice of opacities is important not only for radiation transport but also for the numerical convergence of iterative solvers. Specifically, it is possible that the steep opacity drop associated with hydrogen recombination might cause numerical problems unless properly resolved. Furthermore, velocity gradients in the outflow can cause line smearing, causing differences from the Rosseland-mean opacity typically calculated for static atmospheres. In this work, we use two different opacity prescriptions described below: analytic equations approximating the important sources of opacity and a more realistic opacity table.

\subsubsection{Analytic opacities}
We adopted a simple smooth analytic prescription for opacities similar to what was used in \citet{Pejcha2017} and \citet{Matsumoto2022},
\begin{equation}
\kappa_\text{R}(\rho, T) =\kappa_\text{dust} + \kappa_\text{m} + \kappa_\text{es}\xion +(\kappa_\text{K}^{-1} + \kappa_{\text{H}^{-}}^{-1})^{-1}, \quad \kappa_\text{P} = \kappa_\text{R},
\label{eq:opacity}
\end{equation}
with individual components given by
\begin{align}
\begin{split}
     \kappa_\text{dust} &= \left[\left(10\rho^{1/2}\left(\frac{T}{1500\,\text{K}}\right)^{-10}\right)^{-1} + \kappa_\text{dust,ceiling}^{-1}\right]^{-1},\\
     \kappa_\text{m} &= 2\times 10^{-3}\,\text{cm}^2\,\text{g}^{-1},\\
     \kappa_\text {es} &= 0.34\,\text{cm}^2\,\text{g}^{-1},\\
     \kappa_\text{K} &= 2.75\times 10^{24}\rho T^{-7/2},\\
     \kappa_{H^{-}} &= 1.7\times 10^{-27} \rho^{1/2} T^{7.7},
\end{split}
\label{eq:opacity_detail}
\end{align}
where $\kappa_\text{dust,ceiling} = 3$\,cm$^2$\,g$^{-1}$ and all numerical factors are in cgs units. 

\subsubsection{Tabulated opacities}

We use opacity tables constructed by \citet{tomida13,tomida15}, which combine dust opacities from \citet{semenov03}, low-temperature gas opacities from \citet{ferguson05}, and high-temperature gas opacities from \citet{seaton94}. Details of the construction are given in Appendix~B of \citet{tomida13}. Unlike analytic opacities, these tables provide separate values for Planck and Rosseland mean opacities. The values are tabulated on a dense grid covering $-22 \le \log\rho \le 1$  and $0.3 \le \log T \le 8$ and we use bi-log-linear interpolation to obtain the actual values.

During our work, we found that opacities are the main factor determining the convergence of radiation solver and the length of the timestep. Our simulation setup is particularly challenging in this regard, because we simultaneously consider shocked material with  $\rho \sim 10^{-8}$\,g\,cm$^{-3}$ and $T \sim 10^5$\,K leading to opacities close to the peak values together with wind-like medium with $\rho \sim 10^{-15}$\,g\,cm$^{-3}$ and $T \sim 10^3$\,K and opacities close to the opacity minimum. Consequently, the absorption coefficient product, $\sim \kappa \rho$, easily varies by more than ten orders of magnitude. Furthermore, since $\kappa_\text{P} \gg \kappa_\text{R}$ at $\log \rho \lesssim -5$ and $\log T \gtrsim 3$, another very short timescale, $t_\text{P} \sim (c\kappa_\text{P}\rho)^{-1}$, adversely affects the convergence of the simulation effectively leading to much shorter timestep. In simplified test problems, our code is able to handle situations with $\kappa_\text{P}/\kappa_\text{R} \approx 1000$ \citep{Calderon2021}, but the simulations presented here are significantly more demanding than these test problems. Increasing the value of $t_\text{P}$ by modifying Planck opacities might improve numerical convergence without affecting simulation results if the increased $t_\text{P}$ remains shorter than other relevant timescales.

The analytic prescription (Eq.~[\ref{eq:opacity}]), effectively sets a floor for Rosseland opacity, $\kappa_\text{R,floor} = \kappa_\text{m} = 2\times 10^{-3}$\,cm$^2$\,g$^{-1}$ and a ceiling for Planck opacity, $\kappa_P = \kappa_R$. In order to make successful calculations with tabulated opacities, we had to modify the tables. First, we smoothed the values in $\rho$ and $T$ with a simple Gaussian kernel. This smoothes opacity jumps connected with condensation of various dust species, which are not of interest for our work here. Second, we had to apply a ceiling to Planck opacity, $\kappa_\text{P} = 10\kappa_\text{R}$. We find that making this ceiling less restrictive reduces the timestep almost linearly, preventing completion of simulations within a reasonable time. Third, although we were able to complete several runs without applying floor to $\kappa_\text{R}$ (Table~\ref{tab:params}), most of the attempted runs encountered numerical difficulties near the end of the plateau, which would require reduction of the timestep and again preventing completion of the runs within reasonable time. Consequently, we applied a floor of $\kappa_\text{R,floor} = 10^{-3}$\,cm$^2$\,g$^{-1}$, which is applied before the ceiling on $\kappa_\text{P}$. This floor is lower than the one used in the analytic prescription. We also performed several experiments with lower $\kappa_\text{R,floor}$, which we describe in Section~\ref{app:tests}. Finally, despite these modifications, successful runs required significant increase in the density of the wind-like medium in polar directions compared to the runs with analytic EOS and opacities, but this medium still does not affect the luminosities or other quantities as we demonstrate in Section~\ref{app:tests}.

In Figure~\ref{fig:opacity}, we compare our opacity prescriptions with the original opacity table of \citet{tomida13}. We see that Equation~(\ref{eq:opacity}) (green line) represents very well the dependence of Kramers opacity on density. At low densities, the opacity minimum is shifted to lower temperatures compared to the tabulated values. Applying $\kappa_\text{R,floor}$ (orange lines) becomes important only at lower densities. Ceiling on $\kappa_\text{P}$ significantly lowers the Planck mean opacities, but we illustrate in Section~\ref{app:tests} that the effects are relatively mild.

\subsection{Tests and validation}
\label{app:tests}

\begin{figure*}
\centering
\includegraphics[width=0.5\textwidth]{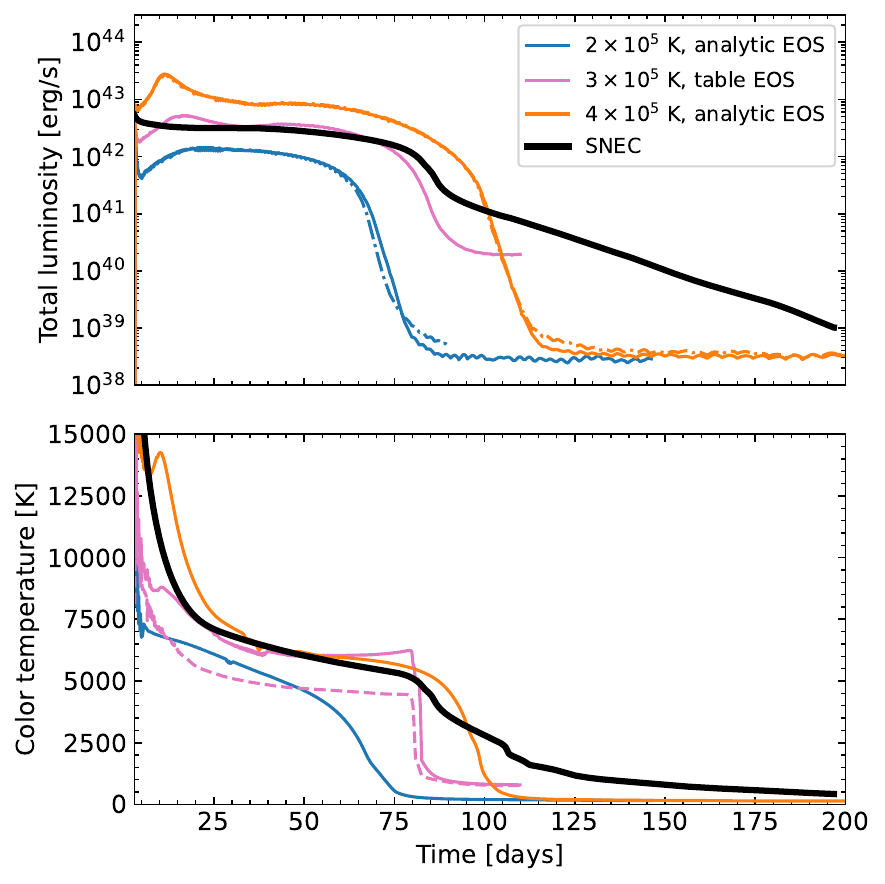}
\caption{Comparison of a theoretical nickel-free Type IIP supernova calculated with SNEC (solid black line) with our simulated runs with $\mej = 10\,\msun$, $E_\text{ej,kin} = 9.8 \times 10^{50}$\,ergs, and three initial ejecta temperatures (blue, pink, and orange solid lines). Upper panel shows luminosities as functions of time. The dash-dotted lines indicate models with the same initial conditions but using a $\Gamma$-law EOS. Lower panel shows the time evolution of Rosseland photospheric temperatures. The dashed pink line shows the time evolution of the Planck photospheric temperature for the model where Rosseland and Planck opacities differ.   \label{fig:iip}}
\end{figure*}

\begin{figure}
\centering
    \includegraphics[width=0.7\textwidth]{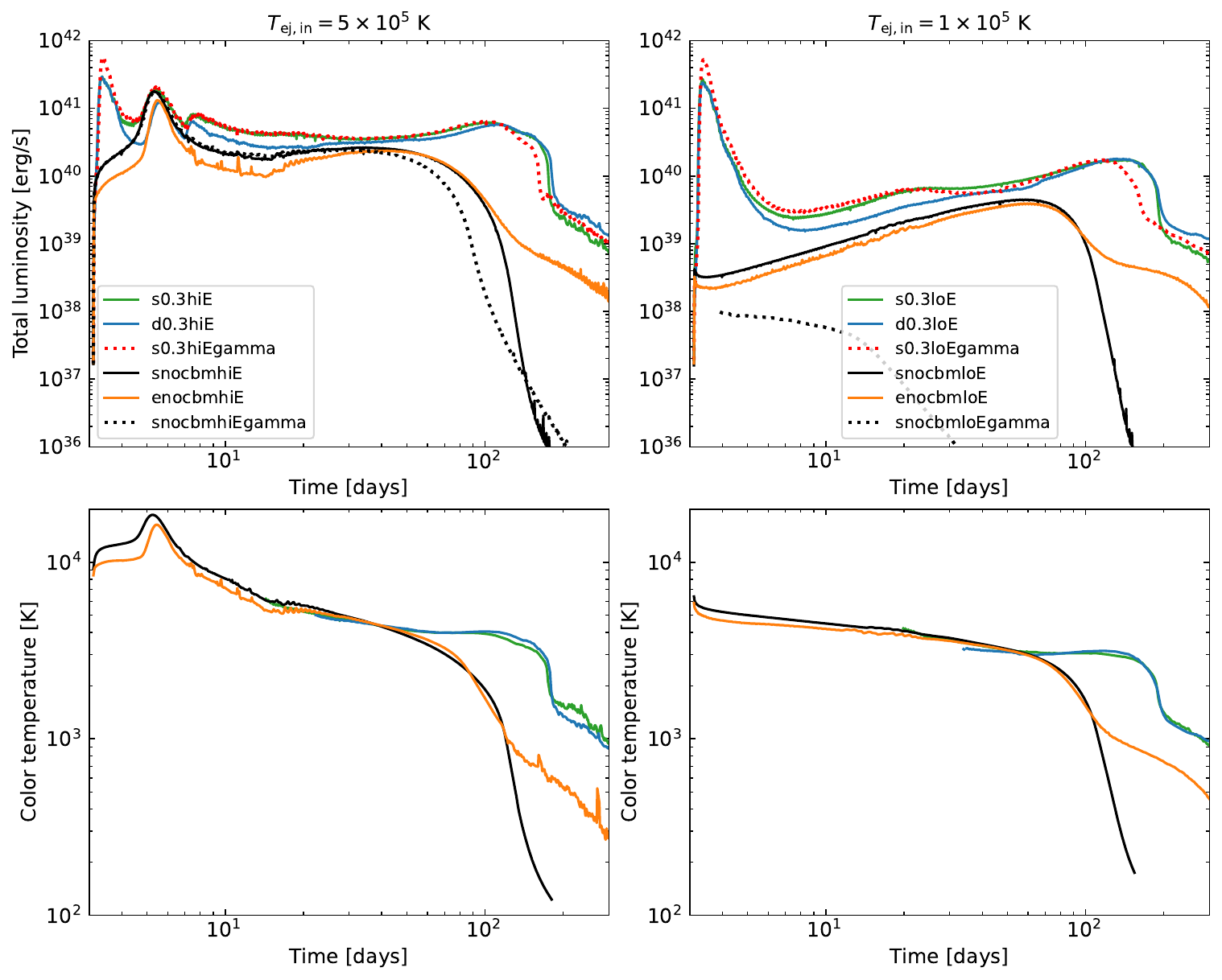}
    \caption{Comparison of the effect of different EOS and $\mdot_\text{wind}$ on luminosity (upper row) and photospheric temperature (lower row) for two different initial ejecta thermal energies (left and right column). Simulation parameters are given in Table~\ref{tab:params}. \label{fig:recomb}}
\end{figure}

\begin{figure}
\centering
    \includegraphics[width=\textwidth]{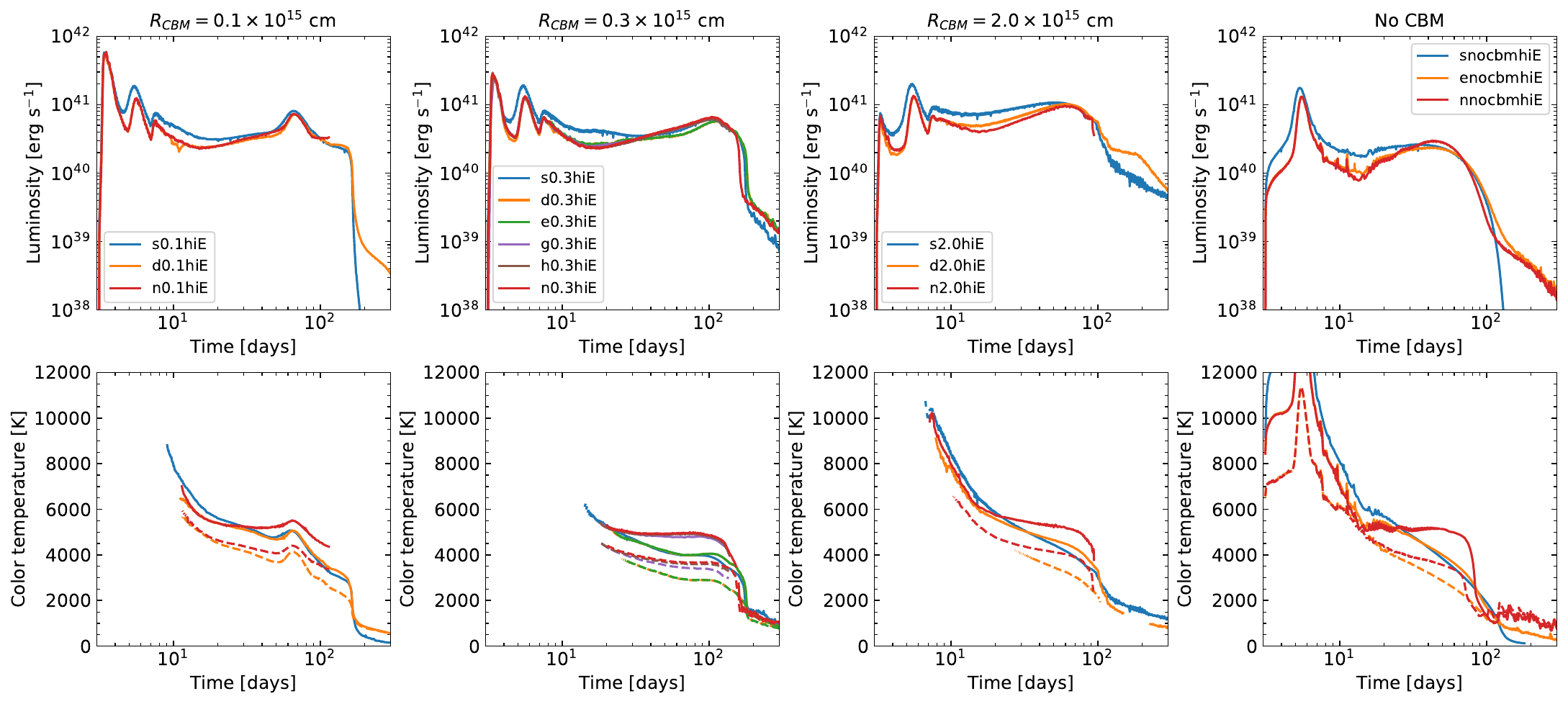}
    \includegraphics[width=\textwidth]{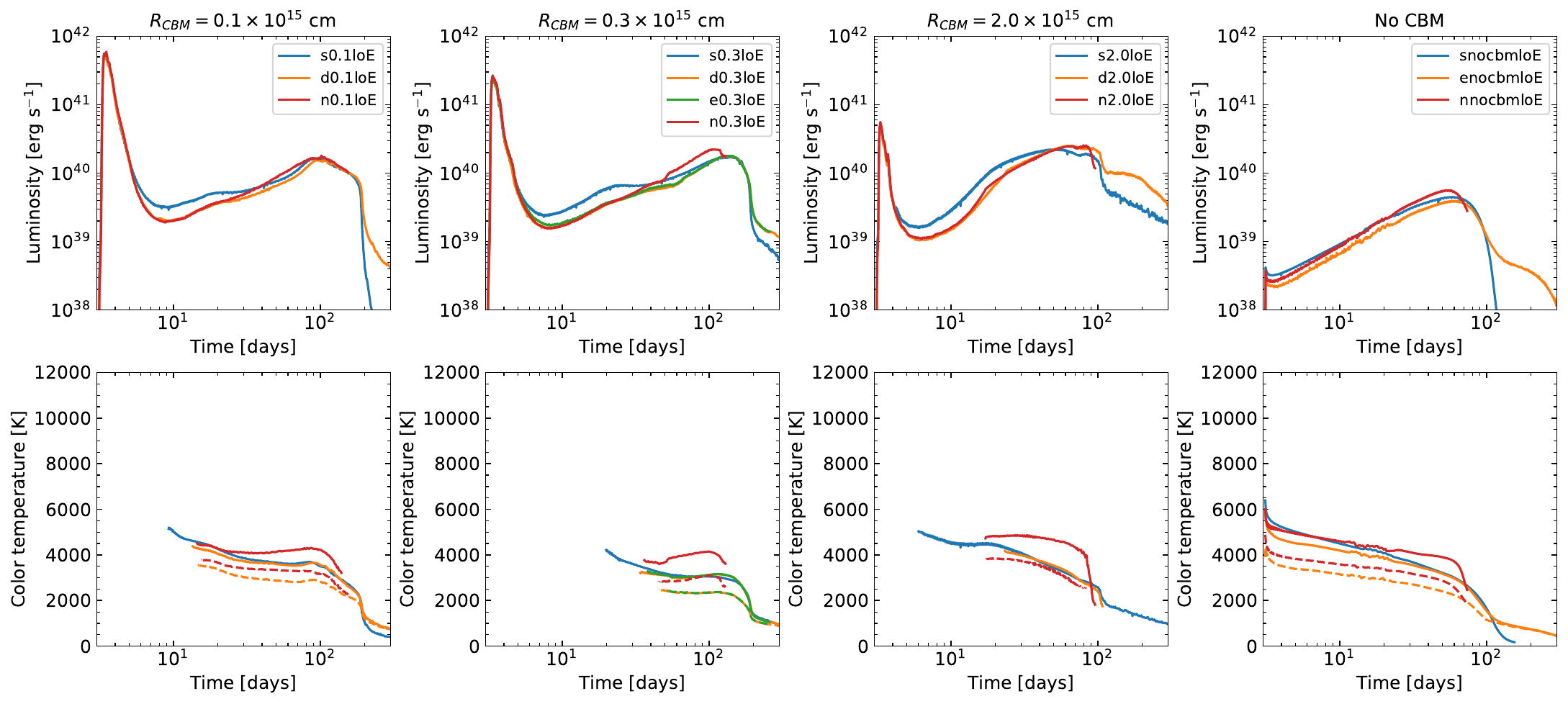}
    \caption{Effect of applying opacity floor and ceiling on the luminosity and photospheric temperature. The upper two rows show results for simulations with $T_\text{ej,in} = 5\times 10^5$\,K (hiE series of models), while the lower two rows show results for $T_\text{ej,in} = 1\times 10^5$\,K (loE series of models). Each column shows results for different $\rcsm$. Model series ``s'' wass calculated with analytic EOS and opacities. All other model series used tabulated EOS and opacities, where series ``d'' presents the default results for runs with CBM. Model series ``e'', ``g'', ``h'', and ``n'' have the same $\mdot_\text{wind}$, which is by a factor of 10 higher than for series ``d'', but this is inconsequential for the light curves. Series ``e'', ``g'', and ``h'' differ by the value of $\kappa_\text{R,floor} = 10^{-3}$, $10^{-4}$, and  $10^{-5}$\,cm$^2$\,g$^{-1}$, respectively. Model series ``n'' does not apply any opacity floor.
    We show also models that had to be terminated before reaching the post-plateau phase due to numerical problems (Table~\ref{tab:params}). \label{fig:lckappa}}
\end{figure}

Here, we shows basic tests and comparisons of our code. First, we performed the same tests for this version of the code as in our previous work \citep{Calderon2021}.
Second, we performed simulations of Type IIP supernovae, which have very well-understood light curves largely controlled by diffusion through the recombining hydrogen envelope \citep[e.g.,][]{arnett80,kasen09}. Our experiment consists of two-dimensional runs similar to those we did for LRNe but with increased mass and kinetic and thermal energy of the ejecta while keeping the form of initial density and temperature profiles (Eqs.~[\ref{eq:rhoej},\ref{eq:temp}]) unchanged. In Figure~\ref{fig:iip}, we compare our results with a one-dimensional simulation with SNEC \citep{Morozova2015} with the default progenitor and without any radioactive nickel.  We do not expect a perfect match, because SNEC explodes the star self-consistently, and we use a simple parameterized ejecta model. Furthermore, SNEC assumes equilibrium between gas and radiation and a relatively high $\kappa_\text{R,floor} = 0.01$\,cm$^2$\,g$^{-1}$ in the envelope. However, we see that our simulations with $\tejin \approx 3\times 10^5$\,K (initial thermal energy at $t=3$\,days equal to $2.9\times 10^{50}$\,erg) are able to match the luminosity and duration of the plateau of the light curve. We also see that the codes disagree on the level of post-plateau luminosity and that for our code the value depends on the used EOS and opacities. As expected, for Type IIP supernovae hydrogen recombination in the EOS does not play any noteworthy role and we obtain effectively the same light curves with a simple $\Gamma$ law EOS. Finally, as we increase $T_\text{ej,in}$, the light curves develop a bump at around $t\approx 10$\,days, which is caused by the not completely realistic initial temperature profile, similarly to what we discuss for LRNe in the main text. We also obtain a relatively good match for the photospheric temperature between the two codes during the plateau phase. The agreement is worse during the first $\sim 20$\,days, where the simulation is influenced by initial conditions, and during the transition to the plateau, where our model nIIP3e5 shows much sharper drop, which is caused by increase in opacity due to dust. Dust formation in our code only manifests through temperature-dependent opacities, which is not entirely realistic in the context of IIP supernova light curves.

In Figure~\ref{fig:recomb}, we show our LRN results for several runs differing only by the combination of EOS and opacities. For the simple $\Gamma$-law EOS, we obtain $\xion$ by solving the Saha equation, but the prescriptions for $e$ and $P$ are different. We see that the difference between $\Gamma$ law and more realistic EOS is minimal for the runs with higher $\lbol$ and during the plateau, because the total amount of recombining energy is subdominant. However, hydrogen recombination is very important for the low-energy run without CBM, where the recombining energy dramatically increases the luminosity. The differences between H-only and H+He EOS are minimal except at late times for models without CBM, where the tabulated H+He EOS leads to a slower decline of luminosity. This implies that helium recombination is not very important for powering the brightest LRNe.

We can also use Figure~\ref{fig:recomb} to assess the effect of $\mdot_\text{wind}$ on the light curves. Higher values of $\mdot_\text{wind}$ will lead to more luminous shocks between the ejecta and the wind, which could potentially affect the light curve especially at times when the overall luminosity is low. But looking at post-plateau evolution, we see that the differences between models with different $\mdot_\text{wind}$ (s0.3hiE, d0.3hiE, e0.3hiE) are minimal implying that ejecta--wind shocks are not biasing our results for luminosity. Furthermore, our analytic model of shock interaction \citep{Metzger2017,pejcha22} predicts that even our highest used value $\mdot_\text{wind}$ will lead to shock powers of only about $10^{37}$\,\ergss.

In Figure~\ref{fig:lckappa}, we investigate the effect of applying opacity floors and ceilings on luminosities and photospheric temperatures. Particularly instructive are models 0.3hiE, where we were able to obtain the longest runs. We see that models with $\kappa_\text{R,floor}=10^{-3}$\,cm$^2$\,g$^{-1}$ (d and e series) give essentially identical results to runs with analytic EOS and analytic opacities with $\kappa_\text{R,floor}=2\times 10^{-3}$\,cm$^2$\,g$^{-1}$ (s series). However, decreasing $\kappa_\text{R,floor}$ to $10^{-4}$\,cm$^2$\,g$^{-1}$ or lower values (g, h, and n series of models) leads to slightly brighter and somewhat shorter plateaus. Similarly, photospheric temperatures are generally higher in these models. This can be understood by realizing that lower $\kappa_\text{R,floor}$ makes the gas more transparent sooner and that we find the photosphere in deeper, hotter regions. These differences should be kept in mind when interpreting the results in the main text.

\begin{figure*}
\centering
\includegraphics[width=0.5\textwidth]{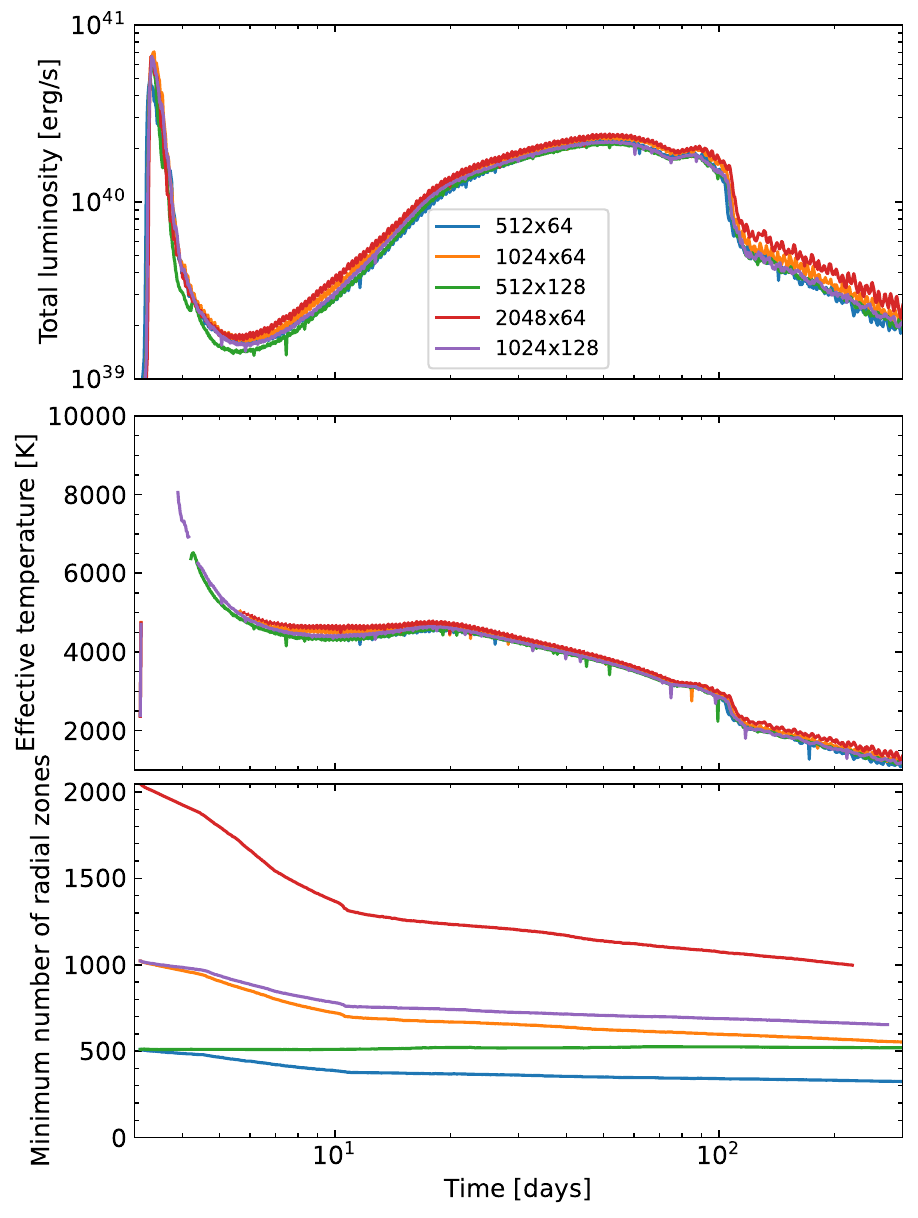}
\caption{Resolution study for CBM runs with $\rcsm = 2.0\times 10^{15}$\,cm and $\tejin = 1 \times 10^5$\,K for different combinations of $N_r \times N_\theta$ given in the legend. \label{fig:resolution}}
\end{figure*}

Finally, in Figure~\ref{fig:resolution}, we show the results of a resolution study for our LRNe runs (s series of models). We see that the differences are very small for the resolutions we have considered. Using even a higher resolution might affect our predictions of $\teff$.

\subsection{Raytracing}
\label{app:ray}

\begin{figure*}
\centering
\includegraphics[width=0.5\textwidth]{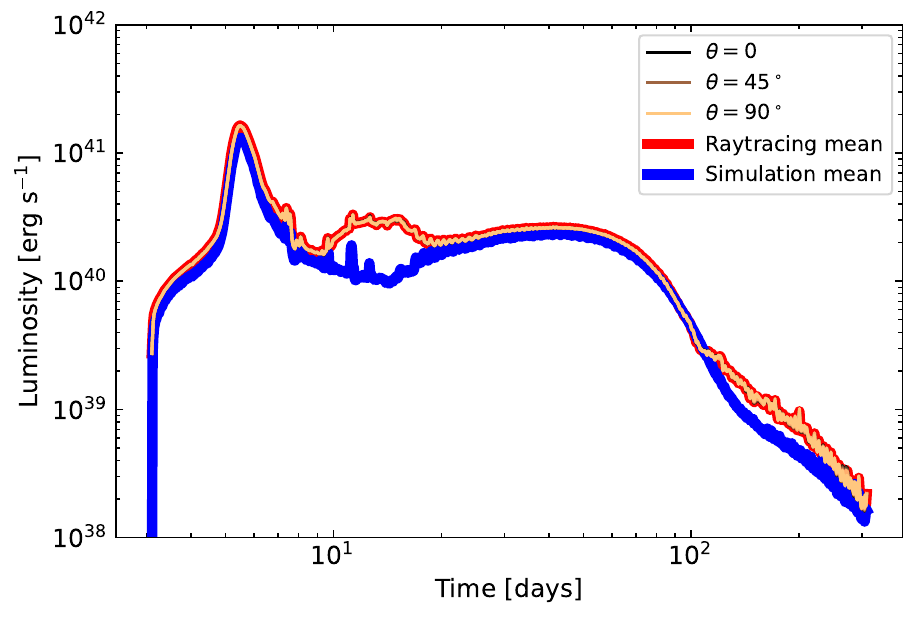}
\caption{Same as the top panel of Figure~\ref{fig:ray} but for spherically-symmetric model without CBM: enocbmhiE.\label{fig:raynocsm}}
\end{figure*}

We constructed a cube composed of $128\times128\times1000$ elements scaled to match the extent of the grid in a given snapshot. We then rotated the cube by nine angles equally spaced between $0^\circ$ (pole on) and $90^\circ$ (edge on) measured from the axis of symmetry. To each cube element, we assigned absorption coefficient $\kappa_\text{P}\rho$ and gas temperature $T$ by log-cubic interpolation in our simulation. We then integrated intensity $I$ along the densely spaced axis using semi-implicit solution radiative transfer equation \citep{Pejcha2017},
\begin{equation}
    I(s+\text{d}s) = e^{-\kappa_\text{P} \rho \text{d}s}I(s) + \left(1-e^{-\kappa_\text{P} \rho \text{d}s}\right)\sigma_\text{SB}T^4,
    \label{eq:ray}
\end{equation}
where $\sigma_\text{SB}$ is the Stefan-Boltzmann constant, and $\text{d}s$ is the distance between elements along the line of integration. This procedure leads to $128\times 128$ images for each angle and each snapshot, which are further integrated to give angle-dependent and angle-averaged luminosities. We used $\kappa_\text{P}$, because Planck opacity couples radiation and matter, and we applied the same floors and ceilings as in our simulations, because this provided the best agreement between the two angle-averaged light curves. We tested our procedure on spherically-symmetric runs without CBM (Fig.~\ref{fig:raynocsm}) and found almost perfect agreement for different orientation angles. We note that we do not expect perfect agreement between raytraced and simulation angle-averaged light curves due to vastly different method in which they were computed. Simulation results are likely more accurate for tracing the overall energy loss due to radiation while raytraced light curves do not suffer from angular diffusion of flux and better capture the viewing angle dependence.

\section{Animated version of figures}
\label{app:movies}

\begin{figure*}
\epsscale{0.8}
\plotone{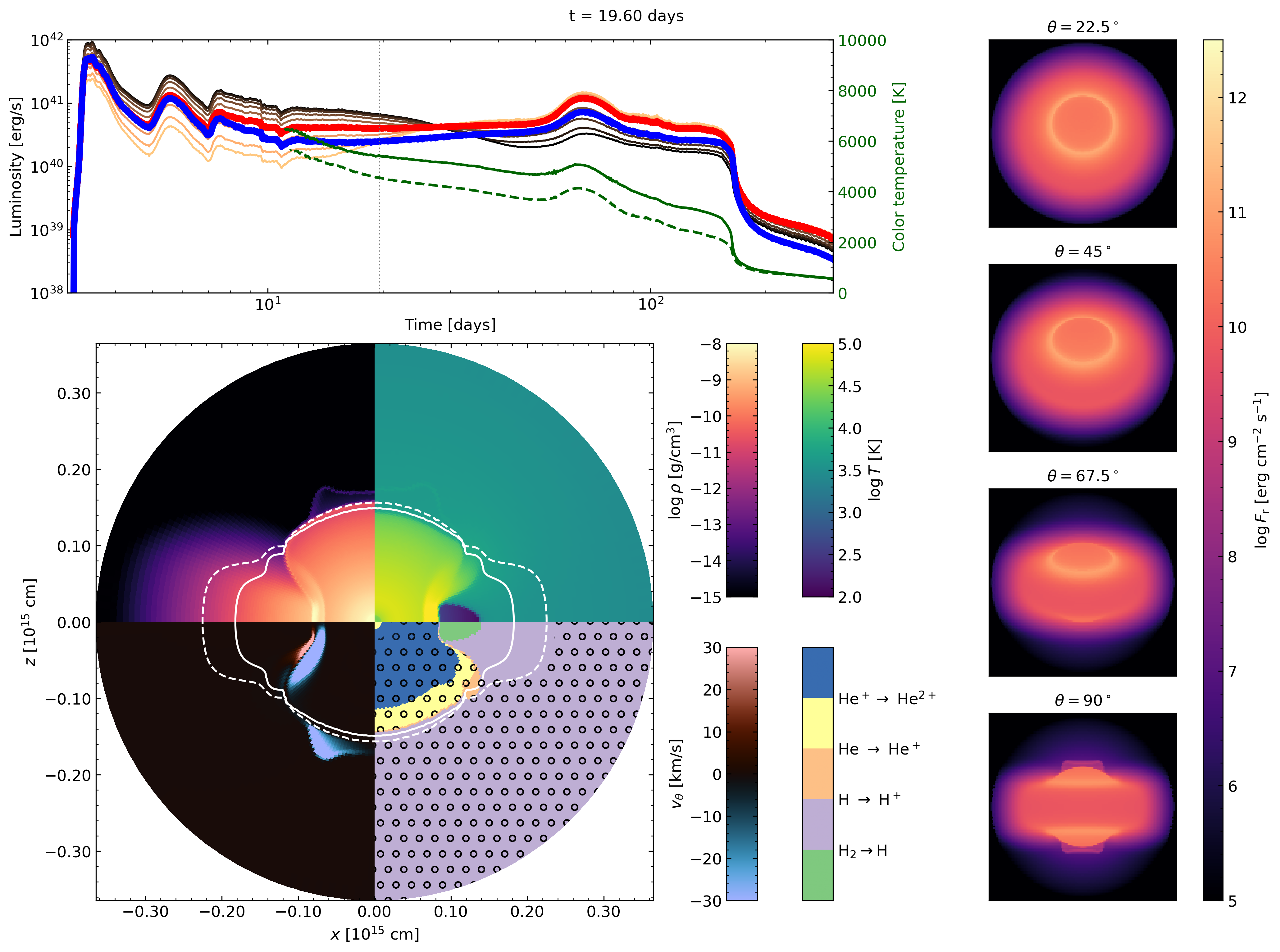}
\caption{Structure and evolution of the simulation d0.1hiE. Top panel shows the time evolution of luminosity from the simulation and raytracing. Meaning of symbols is the same as in Fig.~\ref{fig:ray}. In additon, we show with dark green line evolution of the gas temperature at the photosphere calculated using Rosseland (solid line) and Planck (dashed line) opacities. Left bottom panel shows the instantenous structure of density, gas temperature, tangential velocity, and ionization similarly to Fig.~\ref{fig:overview}. Right panels show raytraced visualizations from several selected angles similarly to Fig.~\ref{fig:ray}. The figure shows one frame of the animation, which lasts approximately 40 seconds.\label{fig:d0.1hiE}}
\end{figure*}

\begin{figure*}
\epsscale{0.8}
\plotone{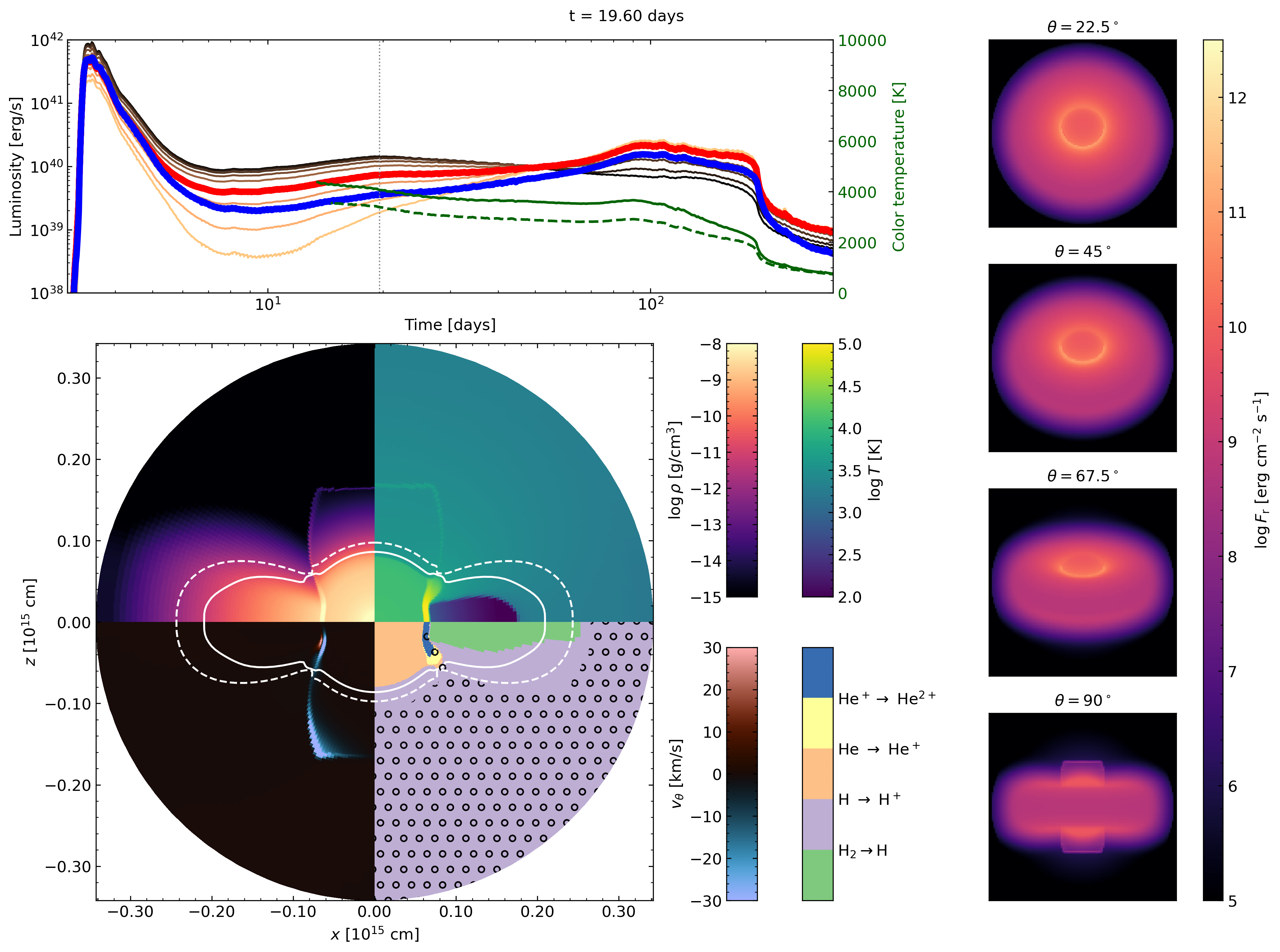}
\caption{Same as Fig.~\ref{fig:d0.1hiE} but for model d0.1loE. \label{fig:d0.1loE}}
\end{figure*}

\begin{figure*}
\epsscale{0.8}
\plotone{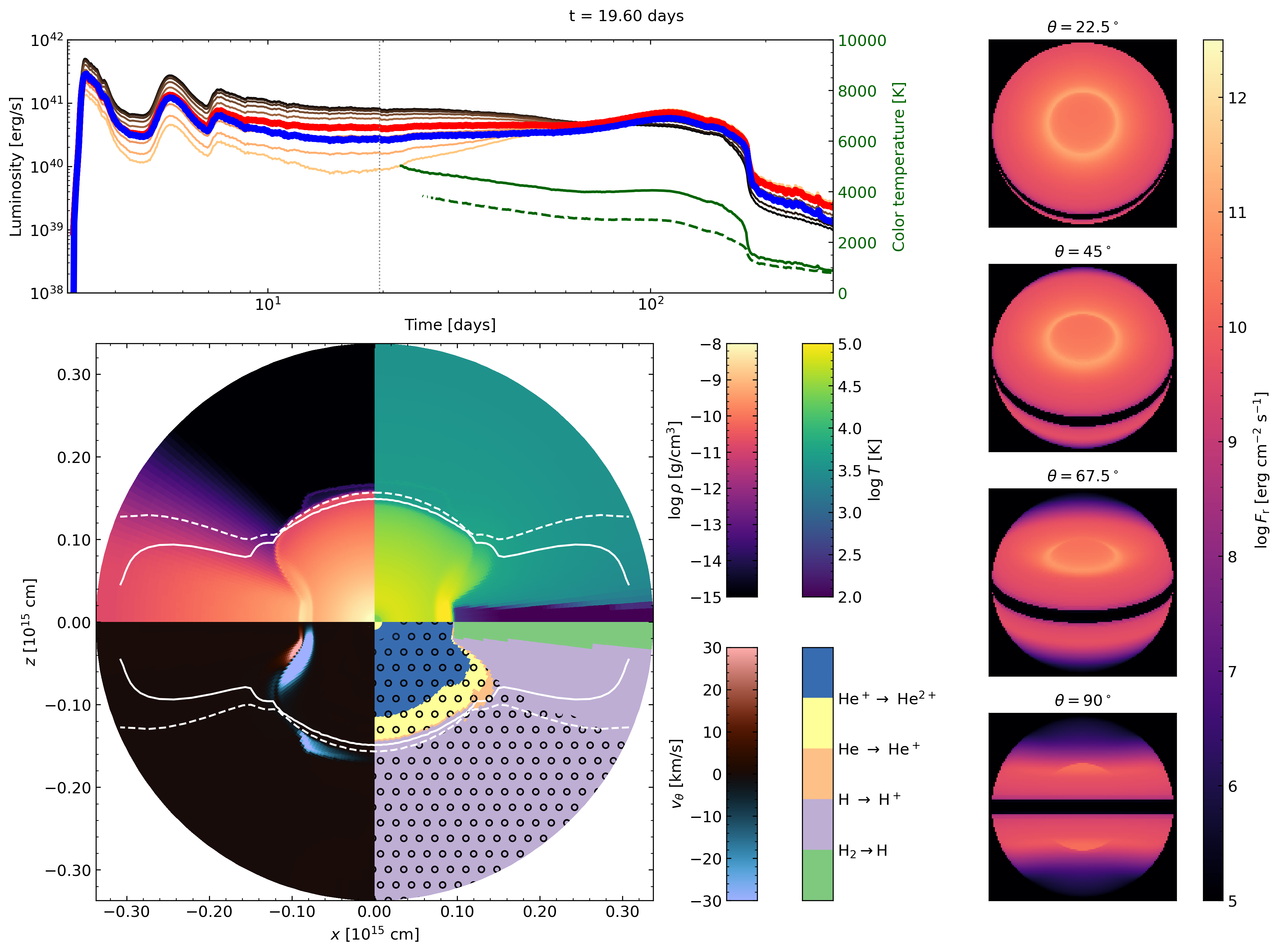}
\caption{Same as Fig.~\ref{fig:d0.1hiE} but for model d0.3hiE. \label{fig:d0.3hiE}}
\end{figure*}

\begin{figure*}
\epsscale{0.8}
\plotone{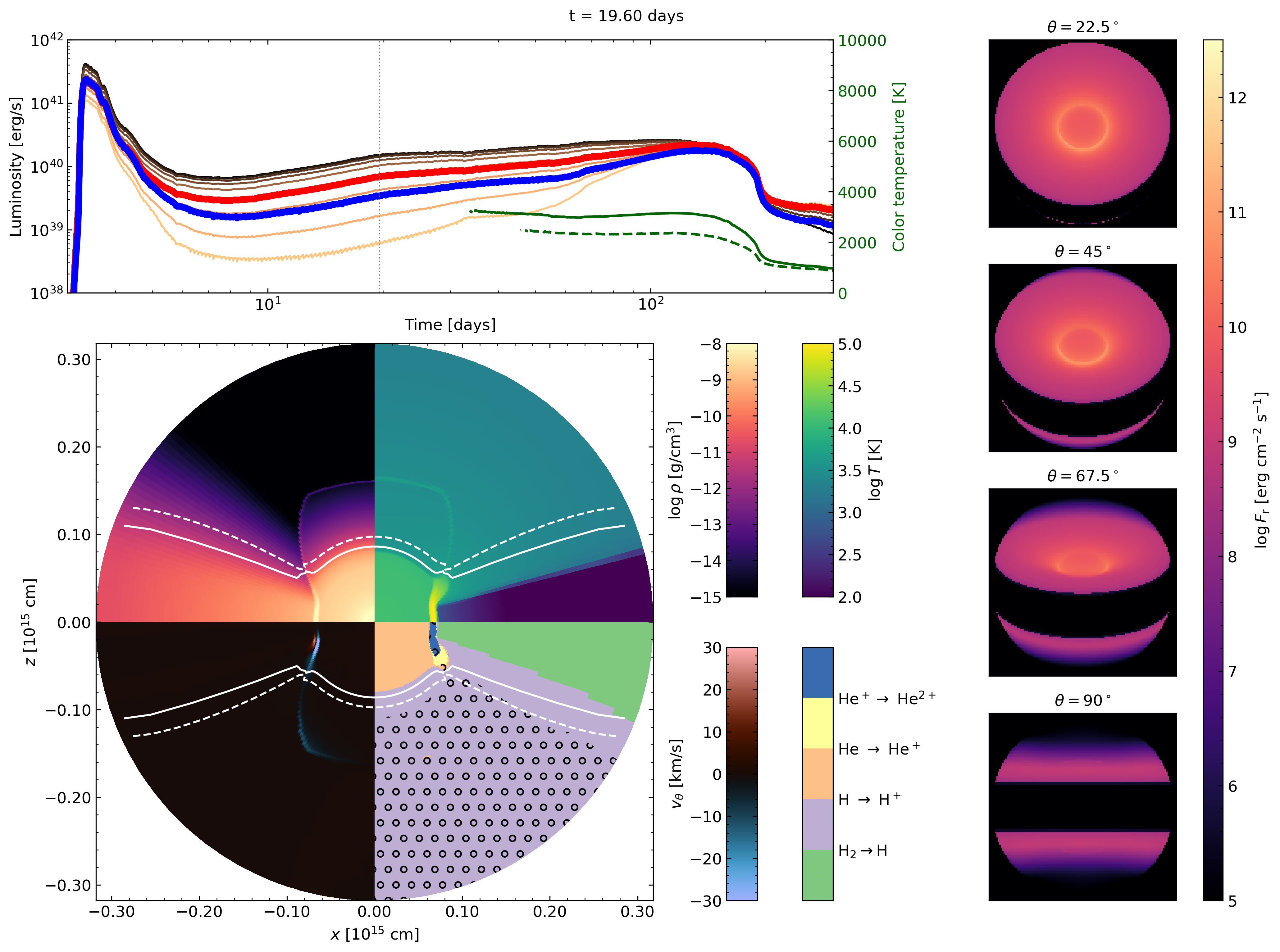}
\caption{Same as Fig.~\ref{fig:d0.1hiE} but for model d0.3loE. \label{fig:d0.3loE}}
\end{figure*}

\begin{figure*}
\epsscale{0.8}
\plotone{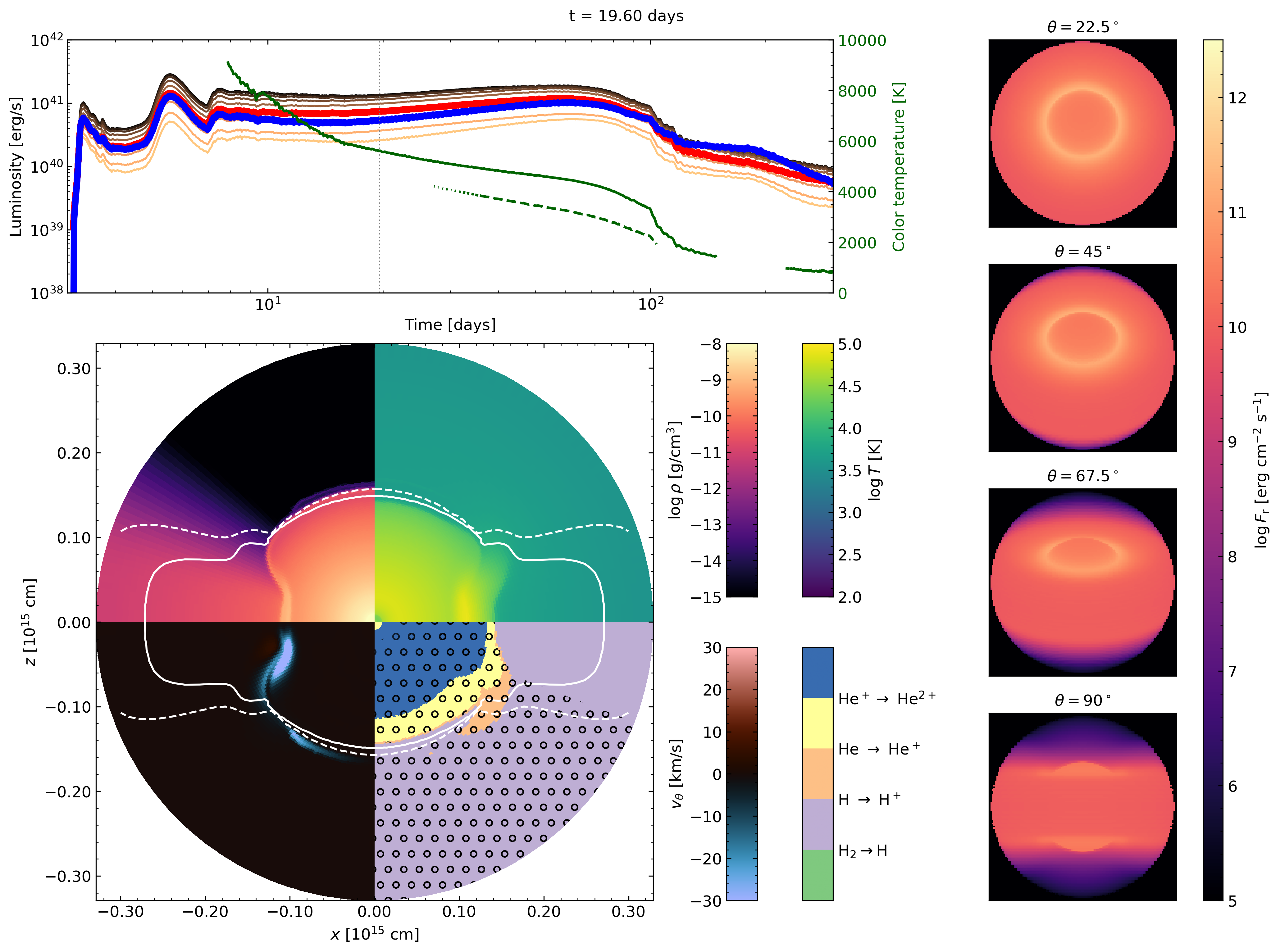}
\caption{Same as Fig.~\ref{fig:d0.1hiE} but for model d2.0hiE. \label{fig:d2.0hiE}}
\end{figure*}

\begin{figure*}
\epsscale{0.8}
\plotone{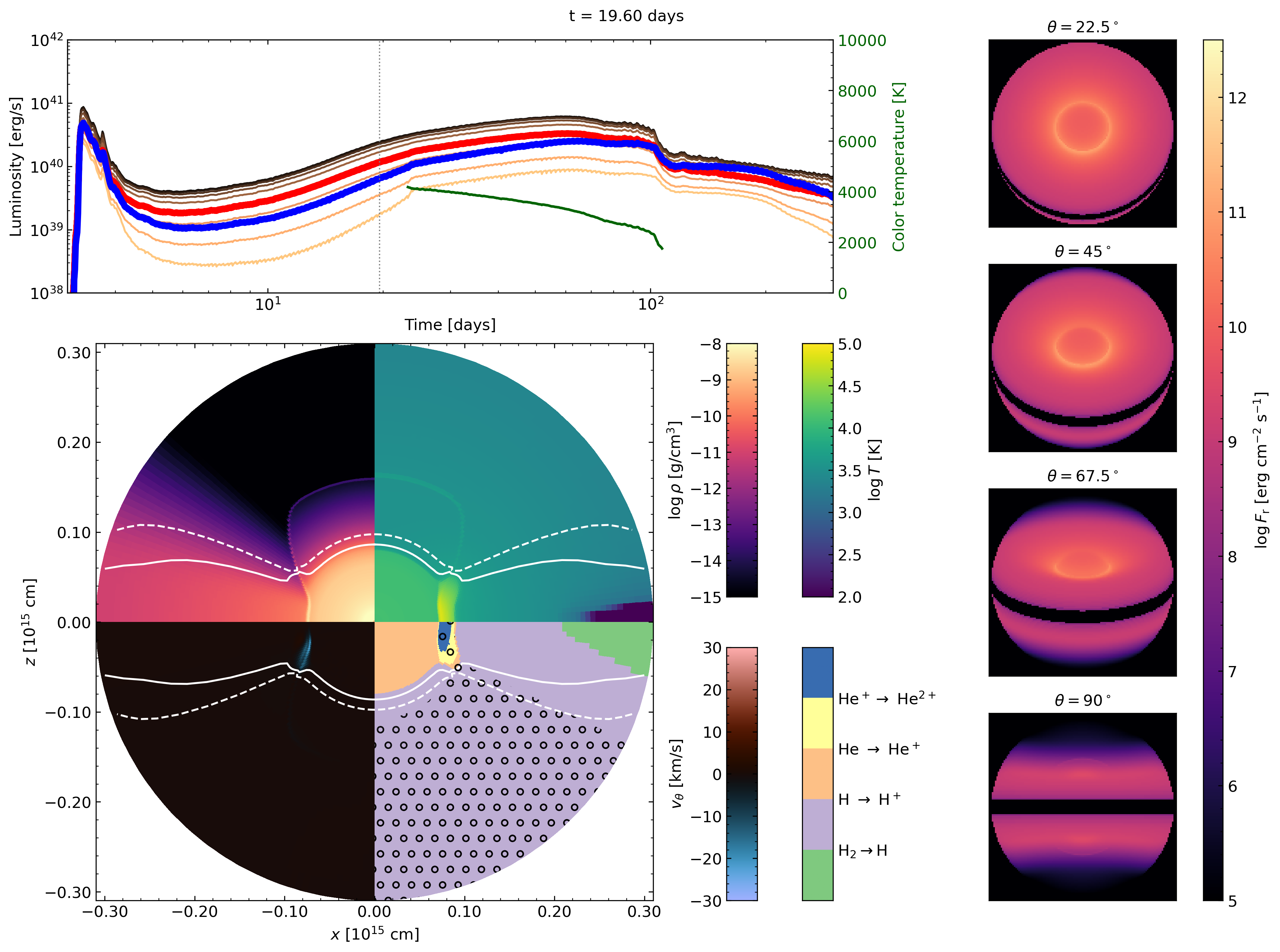}
\caption{Same as Fig.~\ref{fig:d0.1hiE} but for model d2.0loE. \label{fig:d2.0loE}}
\end{figure*}

Here, we present animated versions of figures from the main text for our six CBM simulations, specifically, for models d0.1hiE (Fig.~\ref{fig:d0.1hiE}), d0.1loE (Fig.~\ref{fig:d0.1loE}), d0.3hiE (Fig.~\ref{fig:d0.3hiE}), d0.3loE (Fig.~\ref{fig:d0.3loE}), d2.0hiE (Fig.~\ref{fig:d2.0hiE}), and d2.0loE (Fig.~\ref{fig:d2.0loE}). Movies will become available in the journal once the paper is published.

\bibliography{mybib}{}
\bibliographystyle{aasjournalv7}



\end{document}